\documentclass[onecolumn,english,latin9,bibyear,twocolumn]{aa}
\usepackage[latin9]{inputenc}
\usepackage{textcomp}
\usepackage{amstext}
\usepackage{graphicx}

\makeatletter

\newcommand{\lyxmathsym}[1]{\ifmmode\begingroup\def\b@ld{bold}
  \text{\ifx\math@version\b@ld\bfseries\fi#1}\endgroup\else#1\fi}

\providecommand{\tabularnewline}{\\}

\author{F. Raucq$^{(1)}$, G. Rauw$^{(1)}$, E. Gosset$^{(1)}$, Y. Naz\'e$^{(1)}$, L. Mahy$^{(1)}$, A. Herv\'e$^{(2)}$,
and F. Martins$^{(2)}$}

\institute{$^{(1)}$Department of Astrophysics, Geophysics and Oceanography,
Li\`ege University, Li\`ege, BELGIUM}

\institute{$^{(2)}$LUPM, Montpellier University 2, Montpellier, France}
\authorrunning{F. Raucq et al.}
\titlerunning{Signatures of past mass exchange in HD~149\,404}
\offprints{F.\ Raucq}
\mail{fraucq@doct.ulg.ac.be}

\abstract{Mass and momentum exchanges in close massive binaries play an important role in their evolution, and produce several observational signatures such as asynchronous rotation and altered chemical compositions, that remain after the stars detach again.}
{We investigated these effects for the detached massive O-star binary HD~149\,404 (O7.5\,If + ON9.7\,I, \it  P \rm = 9.81\,days), which is thought to have experienced a past episode of case A Roche-lobe overflow (RLOF).}
{Using phase-resolved spectroscopy, we performed the disentangling of the optical spectra of the two stars. The reconstructed primary and secondary spectra were then analysed with the CMFGEN model atmosphere code to determine stellar parameters, such as the effective temperatures and surface gravities, and to constrain the chemical composition of the components. We complemented the optical study with the study of IUE spectra, which we compare to the synthetic binary spectra. The properties of the stars were compared to evolutionary models.}
{We confirmed a strong overabundance in nitrogen ($\left[\rm N/C\right]\sim150\left[\rm N/C\right]_{\odot}$) for the secondary and a slight nitrogen overabundance ($\left[\rm N/C\right]\sim5\left[\rm N/C\right]_{\odot}$) for the primary star. Comparing the two stars, we found evidence for asynchronous rotation, with a rotational period ratio of $0.50 \pm 0.11$.}
{The hypothesis of a past case A RLOF interaction in HD~149\,404 is most plausible to explain its chemical abundances and rotational asynchronicity. Some of the observed properties, such as the abundance pattern, are clearly a challenge for current case A binary evolution models, however.}% We complement this conclusion by comparing HD~149\,404 with other past case A RLOF systems.}

\usepackage{babel}

\usepackage{babel}

\usepackage{babel}

\usepackage{babel}

\usepackage{babel}

\makeatother

\usepackage{babel}
\begin{document}

\title{\textmd{Observational signatures of past mass-exchange episodes in
massive binaries : The case of HD~149\,404}%
\thanks{Based on observations collected at the European Southern Observatory
(La Silla, Chile) and with the International Ultraviolet Explorer.
The reduced spectra are only available at the CDS via anonymous ftp
to cdsarc.u-strasbg.fr (130.79.128.5) or via http://cdsarc.u-strasbg.fr/viz-bin/qcat?J/A+A/542/A95.%
}}

\author{F.~Raucq\inst{\ref{inst1}}, G. Rauw\inst{\ref{inst1}}, E.
Gosset\inst{\ref{inst1}}%
\thanks{Senior Research Associate F.R.S-FNRS%
}, Y. Nazé\inst{\ref{inst1}}%
\thanks{Research Associate F.R.S-FNRS%
}, L. Mahy\inst{\ref{inst1}}%
\thanks{Postdoctoral Researcher F.R.S-FNRS%
}, A. Hervé\inst{\ref{inst2}}, and F. Martins\inst{\ref{inst3}}}

\institute{Department of Astrophysics, Geophysics and Oceanography, Liège University,
Quartier Agora, Allée du 6 Août 19c, Bat. B5c, B-4000 Liège, Belgium
\label{inst1} \and Astronomical Institute of the Czech Academy of
Sciences, Fri\v{c}ova 298, 251 65 Ond\v{r}ejov, Czech Republic \label{inst2}\and
LUPM, Montpellier University 2, CNRS, place Eugène Bataillon, 34095
Montpellier, France \label{inst3}}

\mail{fraucq@doct.ulg.ac.be}

\maketitle
\makeatother

\section{Introduction}

As shown in recent studies (e.g.\ Sana et al.\ \cite{Sana}), a
large portion of massive stars belongs to binary or higher multiplicity
systems. Whilst this multiplicity permits observationally determining
the minimum masses of the stars through their orbital motion, it also
leads to complications. For instance, the binarity of massive stars
can result in interactions between their stellar winds that produce
observational signatures throughout the electromagnetic spectrum (e.g.
Rauw \cite{Leuven}). Moreover, during the lifetime of a massive binary,
the binarity influences the evolution of the stars in various ways
(e.g. Langer \cite{Langer1}). These evolutionary effects range from
tidally induced rotational mixing (e.g.\ de Mink et al.\ \cite{deMink})
over exchange of matter and angular momentum through a Roche lobe
overflow (RLOF) interaction (e.g.\ Podsiadlowski et al.\ \cite{Podsiadlowski},
de Loore \& Vanbeveren \cite{dLV}, Wellstein et al.\ \cite{Wellstein},
Hurley et al.\ \cite{Hurley}) to the merging of both stars (e.g.\ Podsiadlowski
et al.\ \cite{Podsiadlowski}, Wellstein et al.\ \cite{Wellstein}).
In RLOF interactions, three different situations are distinguished:
case A, for which the RLOF episode occurs when the mass-donor is on
the core-hydrogen burning main-sequence; case B, when the star is
in the hydrogen-shell burning phase; and case C, for which the star
is in the helium-shell burning phase (Kippenhahn \& Weigert \cite{KW},
Vanbeveren et al.\ \cite{VDLVR}). These binary interactions significantly
affect the physical properties of the components and their subsequent
evolution. Despite considerable progresses in theoretical models,
there remain a number of open questions such as the actual efficiency
of accretion (e.g.\ Wellstein et al.\ \cite{Wellstein}, de Mink
et al.\ \cite{deMink1}, Dray \& Tout \cite{DT}). To better understand
this phenomenon, we need in-depth studies of systems undergoing or
having undergone mass exchange.

\medskip{}

In this context, HD~149\,404 is an interesting target. It is a detached,
non-eclipsing O-star binary, member of the Ara OB1 association, located
at a distance of 1.3\,$\pm$ 0.1 kpc (Herbst \& Havlen \cite{Herbst}).
The orbital parameters were revised by Rauw et al. (\cite{Rauw}).
These authors found an orbital period of 9.81 days and a circular
orbit. They also estimated an orbital inclination of 21\textdegree{}
by comparing ``typical'' masses of supergiants with similar spectral
types to those of both components. %The system has a circular orbit with an orbital period of 9.81days. The orbital inclination of the system has been estimated by Rauw et al.() by comparison with typical masses of supergiants of the same spectral type as .
The spectrum of HD~149\,404 displays variable emission lines (He
\textsc{ii} $\lambda$\,4686, H$\alpha$) that are likely indicative
of a wind-wind interaction (Rauw et al.\ \cite{Rauw}, Thaller et
al.\ \cite{Thaller}, Nazé et al.\ \cite{Naze}). One of its components
was assigned an ON spectral type, which suggests a significant nitrogen
enrichment of its atmosphere. %This peculiarity could stem from a past binary interaction.

\medskip{}

In this study, we determine the fundamental parameters of the stars
of this system through several analysis techniques. The data used
in our study are presented in Sect.\ \ref{data}. In Sect.\,\ref{Prelim}
we present the preparatory treatment of our data, which included disentangling
the spectra of the binary system to reconstruct individual spectra
of the binary components, which are needed for the subsequent spectral
analysis. The spectral analyses were made with the non-LTE model atmosphere
code CMFGEN and are presented in Sect.\,\ref{Modelatmosphere}. In
Sect.\,\ref{Conclusions} we discuss the evolutionary state of HD~149\,404
and compare it with similar systems.

\section{Observational data \textmd{\label{data}}}

The optical spectra analysed here are the échelle spectra used by
Rauw et al.\ (\cite{Rauw}) in their study of the orbital solution
and wind interactions of HD~149\,404. These data were collected
in May 1999 and February 2000 with the FEROS and Coralie spectrographs
mounted on the ESO 1.5\,m telescope and 1.2\,m Euler Swiss telescope
at La Silla, respectively. The FEROS and Coralie spectrographs have
resolving powers of 48\,000 and 50\,000 and cover the wavelength
ranges 3650 -- 9200\,\AA{}\ and 3875 -- 6800\,\AA{}, respectively.
Further details on the instrumentation and the data reduction are
given by Rauw et al.\ (\cite{Rauw}). Because of some problems with
the extraction of the FEROS data of May 2000, the normalization of
the latter was more uncertain, and we therefore excluded these data
from the disentengling procedure, and thus from the further analysis.

To complement our optical data, we also retrieved 25 short-wavelength,
high-resolution (SWP) \textit{IUE} spectra of HD~149\,404 from the
\textit{IUE} archives. These spectra were previously described by
Stickland \& Koch (\cite{Stickland}).

Table\,\ref{journal} yields the journal of observations along with
the orbital phases computed according to the ephemerides of Rauw et
al.\ (\cite{Rauw}, their He \textsc{i} $\lambda$\,4471 orbital
solution). The optical spectra provide a good sampling of the orbital
cycle, which is important for the disentangling in Sect.\,\ref{Prelim}.

\begin{table}[htb]
\caption{Journal of the optical (top) and UV (bottom) spectra\label{journal}}

\begin{centering}
\begin{tabular}{ccc}
\hline 
\multicolumn{3}{c}{Optical spectroscopy}\tabularnewline
\hline 
HJD-2\,450\,000  & Inst.  & $\phi$ \tabularnewline
\hline 
1299.796  & FEROS  & 0.24 \tabularnewline
1300.788  & FEROS  & 0.34 \tabularnewline
1301.792  & FEROS  & 0.45 \tabularnewline
1302.783  & FEROS  & 0.55 \tabularnewline
1304.792  & FEROS  & 0.75 \tabularnewline
1578.866  & Coralie  & 0.68 \tabularnewline
1579.879  & Coralie  & 0.78 \tabularnewline
1580.872  & Coralie  & 0.88 \tabularnewline
1581.864  & Coralie  & 0.98 \tabularnewline
1582.869  & Coralie  & 0.09 \tabularnewline
1583.877  & Coralie  & 0.19 \tabularnewline
1584.857  & Coralie  & 0.29 \tabularnewline
1585.845  & Coralie  & 0.39 \tabularnewline
1586.870  & Coralie  & 0.49 \tabularnewline
1587.861  & Coralie  & 0.59 \tabularnewline
1588.886  & Coralie  & 0.70 \tabularnewline
1590.886  & Coralie  & 0.90 \tabularnewline
\end{tabular}
\par\end{centering}

\centering{}%
\begin{tabular}{cccc}
\hline 
\multicolumn{4}{c}{\textit{IUE} SWP spectroscopy}\tabularnewline
\hline 
JD-2\,440\,000  & $\phi$  & JD-2\,440\,000  & $\phi$ \tabularnewline
\hline 
3718.044  & 0.76  & 4450.176  & 0.35 \tabularnewline
3724.588  & 0.42  & 4460.090  & 0.36 \tabularnewline
3776.869  & 0.75  & 9564.333  & 0.42 \tabularnewline
3776.921  & 0.76  & 9565.329  & 0.52 \tabularnewline
3776.966  & 0.76  & 9817.595  & 0.23 \tabularnewline
3793.028  & 0.40  & 9818.630  & 0.33 \tabularnewline
3798.965  & 0.00  & 9840.402  & 0.55 \tabularnewline
3926.010  & 0.95  & 9910.425  & 0.68 \tabularnewline
4001.156  & 0.60  & 9911.396  & 0.78 \tabularnewline
4001.203  & 0.61  & 9912.400  & 0.88 \tabularnewline
4120.089  & 0.72  & 9913.396  & 0.99 \tabularnewline
4129.216  & 0.65  & 9915.391  & 0.19 \tabularnewline
4412.327  & 0.50  &  & \tabularnewline
\hline 
\end{tabular}
\end{table}

\section{Preparatory analysis \textmd{\label{Prelim}}}

\subsection{Spectral disentangling}

The previous determination of the orbital solution of the system performed
by Rauw et al.\ (\cite{Rauw}) allowed us to recover the individual
spectra of both components by separating the normalized spectra of
the binary system. For this purpose, we used our disentangling routine
(Rauw \cite{DSc}), which is based on the method of González \& Levato
(\cite{GL}) and was previously used by Linder et al.\ (\cite{Linder})
and improved by Mahy et al. (\cite{Mahy}). The basic idea behind
this technique is that the most natural way to handle spectra for
radial velocity measurements is to express them as a function of $x=\ln{\lambda}$
instead of as a function of $\lambda$. In this way, the Doppler shift
for a line of rest wavelength $\lambda_{0}$ can be expressed as 
\begin{eqnarray*}
\ln{\lambda} & = & \ln{\left[\lambda_{0}\,\left(1+\frac{v}{c}\right)\right]}\\
 & = & \ln{\lambda_{0}}+\ln{\left(1+\frac{v}{c}\right)}\\
 & \simeq & \ln{\lambda_{0}}+\frac{v}{c}\\
\Rightarrow x & \simeq & x_{0}+\frac{v}{c.}\\
\end{eqnarray*}

Let $A(x)$ and $B(x)$ be the spectra of the primary and secondary
in the heliocentric frame of reference. In the absence of line profile
variability and eclipses, the spectrum observed at any phase $\phi$
can be written as 
\[
S(x)=A(x-\frac{v_{A}(\phi)}{c})+B(x-\frac{v_{B}(\phi)}{c}),
\]
where $v_{A}(\phi)$ and $v_{B}(\phi)$ are the radial velocities
of stars A and B at orbital phase $\phi$.

For a total of $n$ observed spectra, the spectrum of the primary
can be computed as the average of the observed spectra from which
the secondary spectrum shifted by the appropriate radial velocity
(RV) has been subtracted (González \& Levato \cite{GL}) : {\small 
\[
A(x)=\frac{1}{n}\sum_{i=1}^{n}\left[S_{i}\left(x+\frac{v_{A}(\phi_{i})}{c}\right)-B\left(x-\frac{v_{B}(\phi_{i})}{c}+\frac{v_{A}(\phi_{i})}{c}\right)\right]
\]
}And similarly for the secondary: {\small 
\[
B(x)=\frac{1}{n}\sum_{i=1}^{n}\left[S_{i}\left(x+\frac{v_{B}(\phi_{i})}{c}\right)-A\left(x-\frac{v_{A}(\phi_{i})}{c}+\frac{v_{B}(\phi_{i})}{c}\right)\right]
\]
}{\small \par}

Starting from an initial approximation for $B(x)$ (a flat spectrum),
$v_{A}(\phi_{i})$ and $v_{B}(\phi_{i})$, we then iteratively determined
the primary and secondary spectra and their radial velocities (see
González \& Levato \cite{GL}). For the separations we performed here,
we fixed the radial velocities of the binary components to those corresponding
to the orbital solution of Rauw et al.\ (\cite{Rauw}).

\medskip{}

As for any disentangling method, this technique also has its limitations
(González \& Levato \cite{GL}). A reasonable observational sampling
of the orbital cycle is needed because the quality of the results
depends on the radial velocity (RV) ranges covered by these observations.
Another important limitation is that broad spectral features are not
recovered with the same accuracy as narrow ones. This is the case
for the wings of the Balmer lines, but it also holds for the UV features
that arise in the stellar winds, such as P-Cygni profiles. For instance,
for the \textit{IUE} spectra of HD~149\,404, the \ion{Si}{iv}
$\lambda$\,1393-1403 and \ion{N}{iv} $\lambda$\,1718 P-Cygni
lines are very broad and barely vary during the orbital cycle. It
was therefore impossible to properly separate the primary and secondary
contributions for these lines. In practice, features that are wider
than a few times the RV amplitude are barely recovered. Moreover,
small residual errors in the normalization of the input spectra can
lead to oscillations of the continuum in the resulting separated spectra
on wavelength scales of several dozen \AA{}. Finally, spectral disentangling
works on continuum-normalized spectra and does not yield the brightness
ratio of the stars, which must be determined by other techniques (see
below).

Finally, in the specific case of HD~149\,404, several emission lines
(H$\alpha$ and He II $\lambda$ 4686, for example) do not seem to
arise in the atmosphere of either component, but are formed at least
partly in the wind-wind interaction zone (Rauw et al.\ \cite{Rauw},
Thaller et al.\ \cite{Thaller}, Nazé et al.\ \cite{Naze}). As
a result, the disentangling code is unable to meaningfully reconstruct
these lines in the spectra of the components. We need to keep this
in mind when analysing the reconstructed spectra.

\subsection{Spectral types}

Based on the reconstructed individual line spectra of the primary
and secondary components, we determined the spectral types of the
stars by measuring the equivalent width ratio of the spectral lines
He\,\textsc{i} $\lambda$\,4471 and He\,\textsc{ii} $\lambda$\,4542
on the one hand, and Si\,\textsc{iv} $\lambda$\,4088 and He\,\textsc{i}
$\lambda$\,4143, on the other hand. We applied Conti's quantitative
classification criteria for O-type stars (Conti \& Alschuler \cite{CA},
Conti \& Frost \cite{CF}, see also van der Hucht \cite{vdHucht})
for both spectral types and luminosity classes to determine that the
primary star is an O7.5\,If star, while the secondary is an ON9.7\,I%
\footnote{In the literature on interacting binaries, the terminologies ``primary''
and ``secondary'' usually designate the initially more massive star
and less massive star, respectively. However, in accordance with previous
studies of the HD~149\,404 system, in the present paper the terms
``primary'' and ``secondary'' star refer to the currently more
massive and less massive component, respectively.%
}. Since our classification is based on the separated spectra, it is
less sensitive to a possible phase-dependence of the line strengths
and should thus be more robust than a classification based on spectra
that are only collected near quadrature phase, as done by Rauw et
al.\ (\cite{Rauw}). However, in this particular case, we find an
excellent agreement with the classification (O7.5\,I(f) + ON9.7\,I)
proposed by these authors.

\subsection{Brightness ratio \label{brightness}}

Spectral disentangling yields the strength of the lines in the primary
and secondary stars relative to the combined continuum, but does not
allow establishing the relative strengths of the continua. To further
analyse the separated spectra, we therefore need to first establish
the brightness ratio of the stars.

To estimate the optical brightness ratio of the components of HD~149\,404,
we measured the equivalent widths of a number of spectral lines on
the separated spectra of the primary and secondary, but referring
to the combined continuum of the two stars. The results are listed
in Table\,\ref{EW} along with the mean equivalent widths of the
same lines in synthetic spectra of stars of the same spectral type,
computed with the non-LTE model atmosphere code CMFGEN (Hillier \&
Miller \cite{HM}), which we describe in Sect.\,\ref{CMFGENcode}.
For comparison, we also list the mean equivalent widths of the same
lines in spectra of stars of the same spectral type as evaluated from
the measurements of Conti (\cite{Conti1,Conti2}) and Conti \& Alschuler
(\cite{CA}).

\begin{table*}[thb]
\caption{Brightness ratio determination from the dilution of prominent lines}

\begin{centering}
\begin{tabular}{l|ccccccc}
\hline 
Line  & \multicolumn{6}{c}{Equivalent Width (\AA{})} & \multicolumn{1}{c}{$l{}_{1}/l_{2}$}\tabularnewline
\hline 
 & \multicolumn{2}{c}{Observations} & \multicolumn{2}{c}{Synthetic spectra} & \multicolumn{2}{c}{Conti (\cite{Conti1,Conti2}) } & \tabularnewline
\hline 
 & Primary  & Secondary  & O7.5  & O9.7  & O7.5  & O9.5  & \tabularnewline
\hline 
\hline 
He\,\textsc{i} $\lambda$\,4026  & 0.19  & 0.34  & 0.72  & 0.74  & 0.65  & 0.80  & 0.57\tabularnewline
Si\,\textsc{iv }$\lambda$\,4089  & 0.12  & 0.34  & 0.32  & 0.56  & 0.25  & 0.53  & 0.61\tabularnewline
Si\,\textsc{iv }$\lambda$\,4116  & 0.06  & 0.24  & 0.11  & 0.29  & 0.12  & 0.37  & 0.66\tabularnewline
He\,\textsc{ii} $\lambda$\,4200  & 0.11  & 0.08  & 0.47  & 0.21  & 0.42  & 0.23  & 0.61\tabularnewline
H$\gamma$  & 0.45  & 0.50  & 1.77  & 1.74  & 1.88  & 2.09  & 0.88\tabularnewline
He\,\textsc{i} $\lambda$\,4471  & 0.17  & 0.37  & 0.55  & 0.69  & 0.71  & 0.97  & 0.58\tabularnewline
He\,\textsc{ii} $\lambda$\,4542  & 0.17  & 0.07  & 0.53  & 0.21  & 0.64  & 0.25  & 0.96\tabularnewline
\hline 
\end{tabular}
\par\end{centering}

\tablefoot{The measured EWs are compared with values for the same
lines in synthetic spectra of the same spectral type and in the compilation
of measurements from the literature. The last column yields the brightness
ratio for each line considered, using the synthetic spectra EWs. \label{EW}} 
\end{table*}

The brightness ratio of the two stars can then be evaluated from

\begin{center}
$\frac{l_{1}}{l_{2}}=(\frac{EW_{1}}{EW_{2}})_{obs}(\frac{EW_{O9.7}}{EW_{O7.5}})_{mean}$
. 
\par\end{center}

By combining our measurements with those from synthetic spectra, we
derive an optical brightness ratio of $0.70\pm0.12$ in this way,
which is slightly lower than the value found by Rauw et al.\ (\cite{Rauw}),
who inferred $0.90\pm0.16$. As a consistency check, we also determined
the brightness ratio by comparison with Conti's measurements and obtained
$0.79\pm0.13$. The difference between the results obtained based
on synthetic spectra versus Conti's measurements arises at least partially
because Conti (\cite{Conti1,Conti2}) did not take into account the
spectral type O9.7, but instead considered an O9.5 spectral type,
consisting of both classes of stars currently known as O9.5 and O9.7.

Reed (\cite{Reed}) reported a mean $V$ magnitude of 5.48 $\pm$
0.02 and $B-V$ colour of $0.39\pm0.02$ for the system. Since the
intrinsic $(B-V)_{0}$ of an O7.5 or O9.5 star is $-0.26$ (Martins
\& Plez \cite{MP}), the extinction $A_{V}$ can be calculated as
being $2.02\pm0.06$, assuming $R_{V}=3.1$. Assuming the 1.3\,$\pm$\,0.1
kpc distance (Wolk et al.\ \cite{Wolk}), we infer an absolute $M_{V}=-7.11\pm0.07$
for the binary. A brightness ratio of $l{}_{1}/l_{2}=0.70\pm0.12$
then yields individual absolute magnitudes of $M_{V}^{P}=-6.15\pm0.13$
and $M_{V}^{S}=-6.53\pm0.10$.

The separated continuum-normalized primary and secondary optical spectra
are shown in Fig.\ \ref{fig1}.

\section{Spectral analysis \textmd{\label{Modelatmosphere}}}

\subsection{Rotational velocities and macroturbulence}

After applying a Fourier transform method (Simón-Díaz \& Herrero \cite{Simon-Diaz},
Gray \cite{Gray}) to the profiles of the He\,\textsc{i} $\lambda\lambda$\,4026,
4922, 5016, O\,\textsc{iii} $\lambda$\,5592, and C\,\textsc{iv}
$\lambda$\, 5812 lines of the separated spectra, we determined the
projected rotational velocities $v\,\sin{i}$ of the stars of the
system. We selected these lines because they are very well isolated
in the spectra and are therefore expected to be free of blends. The
results are presented in Table\,\ref{vsini}. The mean $v\,\sin{i}$
of the primary star is $(93\pm8)$\,km\,s$^{-1}$, while that of
the secondary is $(63\pm8)$\,km\,s$^{-1}$. %These values are somwhat lower than those inferred by Howarth et al.() from IUE spectra cross-correlated with the spectrum of Sco. Indeed, the latter authors inferred 100 and 91kms for the primary and secondary respectively.

\begin{table}[h]
\caption{Projected rotational velocities ($v\,\sin{i}$ in km\,s$^{-1}$)
of the components of HD~149\,404}

\label{vsini}

\centering{}%
\begin{tabular}{c|cc}
\hline 
Line  & Primary  & Secondary\tabularnewline
\hline 
\hline 
He\,\textsc{i} $\lambda$\,4026  & 97  & 75\tabularnewline
He\,\textsc{i} $\lambda$\,4922  & 94  & 64\tabularnewline
He\,\textsc{i} $\lambda$\,5016  & 105  & 58\tabularnewline
O\,\textsc{iii} $\lambda$\,5592  & 87  & 55\tabularnewline
C\,\textsc{iv} $\lambda$\,5812  & 82  & -\tabularnewline
\hline 
Mean value  & 93 $\pm$ 8  & 63 $\pm$ 8\tabularnewline
\hline 
\end{tabular}
\end{table}

Macroturbulence is defined as a non-thermal motion in the stellar
atmosphere in which the size of the turbulent cell is greater than
the mean free-path of the photons. The main effect of the macroturbulence
is an additional broadening of the spectral lines. This effect has
received much attention over recent years (e.g.\ Simón-D\'{i}az
\& Herrero \cite{Simon-Diaz-1}). A first approximation of the macroturbulence
velocities is obtained by applying the radial-tangential anisotropic
macroturbulent broadening formulation of Gray (\cite{Gray}) on the
spectra after including rotational velocity broadening. We used the
auxiliary program MACTURB of the stellar spectral synthesis program
SPECTRUM v2.76 developed by Gray (\cite{macturb}). We applied this
technique to the lines \ion{He}{i} $\lambda\lambda$ 4026, 4471,
and 5016, and \ion{He}{ii} $\lambda$ 4542. In this way, we obtained
macroturbulence velocities of 70 and 80\,km\,s$^{-1}$ for the primary
and secondary stars, respectively. Although this is quite high, these
numbers are consistent with measurements of Simón-D\'{i}az \& Herrero
(\cite{Simon-Diaz-1}) that were made on stars of similar spectral
types.

\medskip{}

Both rotational and macroturbulence velocities were applied on the
synthetic spectra (see Sect.\,\ref{CMFGENcode}) before comparing
the latter with the separated spectra.

\subsection{CMFGEN code \label{CMFGENcode}}

To determine the fundamental properties of the components of HD~149\,404,
we used the non-LTE model atmosphere code CMFGEN (Hillier \& Miller
\cite{HM}). This code solves the equations of radiative transfer
and statistical equilibrium in the co-moving frame for plane-parallel
or spherical geometries for Wolf-Rayet stars, O stars, luminous blue
variables, and supernovae. CMFGEN furthermore accounts for line blanketing
and its effect on the energy distribution. The hydrodynamical structure
of the stellar atmosphere is specified as an input to CMFGEN. For
the stellar wind, a $\beta$ law is used to describe the velocity
law. In the resolution of the equations of statistical equilibrium,
a super-level approach is adopted. The following chemical elements
and their ions are included in the calculations: H, He, C, N, O, Ne,
Mg, Al, Si, S, Ca, Fe, and Ni. The solution of the equations of statistical
equilibrium is used to compute a new photospheric structure, which
is then connected to the same $\beta$ wind velocity law. The radiative
transfer equations are solved based on the structure of the atmosphere,
with a microturbulent velocity varying linearly with velocity from
10 km s$^{-1}$ in the photosphere to $0.1$ \texttimes{} v$_{\infty}$
at the outer boundary, and generated synthetic spectra are compared
to the separated spectra.

\medskip{}

As a first approximation, we assumed that gravity, stellar mass, radius,
and luminosity are either assumed equal to typical values of stars
of the same spectral type (Martins et al.\ \cite{Martins2}) or adopted
from the study of Rauw et al.\ (\cite{Rauw}). The mass-loss rates
and $\beta$ parameters were taken from Muijres et al.\ (\cite{Muijres})
for the spectral types of both stars, whilst the wind terminal velocity
was set equal to 2450\,km\,s$^{-1}$ (Howarth et al.\ \cite{Howarth})
for the two components.

\medskip{}

The relevant parameters were then adjusted by an iterative process
because each adjustment of a given parameter leads to some modifications
in the value of others. This process and the results we obtained are
presented in Subsects. \ref{Method} and \ref{Results}.

\subsection{Method\label{Method}}

The first step was to adjust the temperature in the stellar models.
The temperatures were mostly determined through the relative strengths
of the He\,\textsc{i} $\lambda$\,4471 and He\,\textsc{ii} $\lambda$\,4542
lines (Martins \cite{Martins}). The final values are 34\,000 and
28\,000 K for the primary and secondary stars. Both results agree
reasonably well with the temperatures expected for supergiants of
these spectral types (Martins et al.\ \cite{Martins2}).

Subsequently, the natural steps to follow would be adjusting the surface
gravities, the mass-loss rate, and the clumping factor, but the binarity
of the studied system causes some problems. Indeed, since the wings
of broad lines are not properly restored by the disentangling, it
is impossible to use the Balmer lines to constrain the surface gravity.
And since the H$\alpha$ and H$\beta$ lines are in addition strongly
affected by extra emission from the wind-wind interaction (Rauw et
al.\ \cite{Rauw}, Thaller et al.\ \cite{Thaller}, Nazé et al.\ \cite{Naze}),
they cannot be used as diagnostics of the mass-loss rate and clumping
factor. The mass-loss rates were therefore fixed to the values tabulated
by Muijres et al.\ (\cite{Muijres}) for stars of the same spectral
type (see their Table\,1, method A). The clumping formalism used
in the CMFGEN model is

\begin{center}
$f(r)=f_{1}+(1\lyxmathsym{\textminus}f_{1})e^{(\lyxmathsym{\textminus}\frac{V(r)}{f_{2}})}$ 
\par\end{center}

\noindent with a filling factor $f_{1}$ of 0.1, a clumping velocity
factor $f_{2}$ of 100 kms$^{-1}$ , and V(r) the velocity of the
wind.

\medskip{}

We used an iterative process to constrain the luminosities and surface
gravities. The first estimate of $\log(g)$ was taken from Martins
et al.\ (\cite{Martins2}) following the spectral types of the stars.
Combined with our determination of the effective temperatures, we
then inferred the bolometric corrections (Lanz \& Hubeny \cite{LH}).
This in turn yielded the individual bolometric luminosities, using
the absolute $V$ magnitudes of the components derived in Sect.\,\ref{brightness}.

These bolometric luminosities and the effective temperatures were
used to compute the ratio of the stellar radii $\frac{R_{P}}{R_{S}}$.
Together with the assumed gravities, this ratio yields the spectroscopic
mass ratio $\frac{M_{P}}{M_{S}}$ , which was then compared to the
dynamical mass ratio inferred from the orbital solution (Rauw et al.
\cite{Rauw}). The difference in these mass ratios was used to produce
a new estimate of the surface gravities, and this iterative process
was repeated until the spectroscopic and dynamical mass ratios agreed
with each other and the CMFGEN synthetic spectra produced for the
new gravity values matched the observations as well as possible.

\medskip{}

After the fundamental stellar parameters were established or fixed,
we investigated the CNO abundances through the strengths of the associated
lines. We performed a normalized $\chi^{2}$ analysis to determine
the best fit to selected lines (Martins et al. \cite{Martins3}).
The normalization consists in dividing the $\chi^{2}$ by its value
at minimum, $\chi\text{\texttwosuperior}{}_{min}$. As a 1$\sigma$
uncertainty on the abundances, we then considered abundances up to
a $\chi^{2}$ of 2.0, that is, an approximation for 1 over $\chi\text{\texttwosuperior}{}_{min}$,
as suggested by Martins et al. (\cite{Martins3}). We used the \ion{C}{iii}
$\lambda\lambda$\,4068-70, \ion{C}{ii} $\lambda$\,4267, \ion{C}{iii}
$\lambda$\,5826, \ion{N}{iii} $\lambda$\,4004, \ion{N}{iii}
$\lambda\lambda$\,4511-15-18-24, \ion{N}{ii} $\lambda$\,5026,
\ion{O}{iii} $\lambda$\,5508, and \ion{O}{iii} $\lambda$\,5592
lines to adjust the C, N, and O abundances of the primary star. For
the secondary star, the same lines were used for the N and O abundances,
but the only line used for the C abundance was \ion{C}{iii} $\lambda\lambda$\,4068-70.
Two methods were used to determine the $\chi^{2}$ of each line. First,
we compared the depth of the observed and synthetic line cores in
the calculation. Second, we determined the mean $\chi^{2}$ for every
independent point of the full profile of each considered line. The
two methods gave similar results. After testing the influence of the
errors on projected rotational velocities on the determination of
the surface abundances, we found that their effect was small compared
to the 1$\sigma$ uncertainties on the measured abundances as determined
here above. We therefore neglected them.

\subsection{Results\label{Results}}

\begin{figure*}[!t]
\includegraphics[scale=0.48]{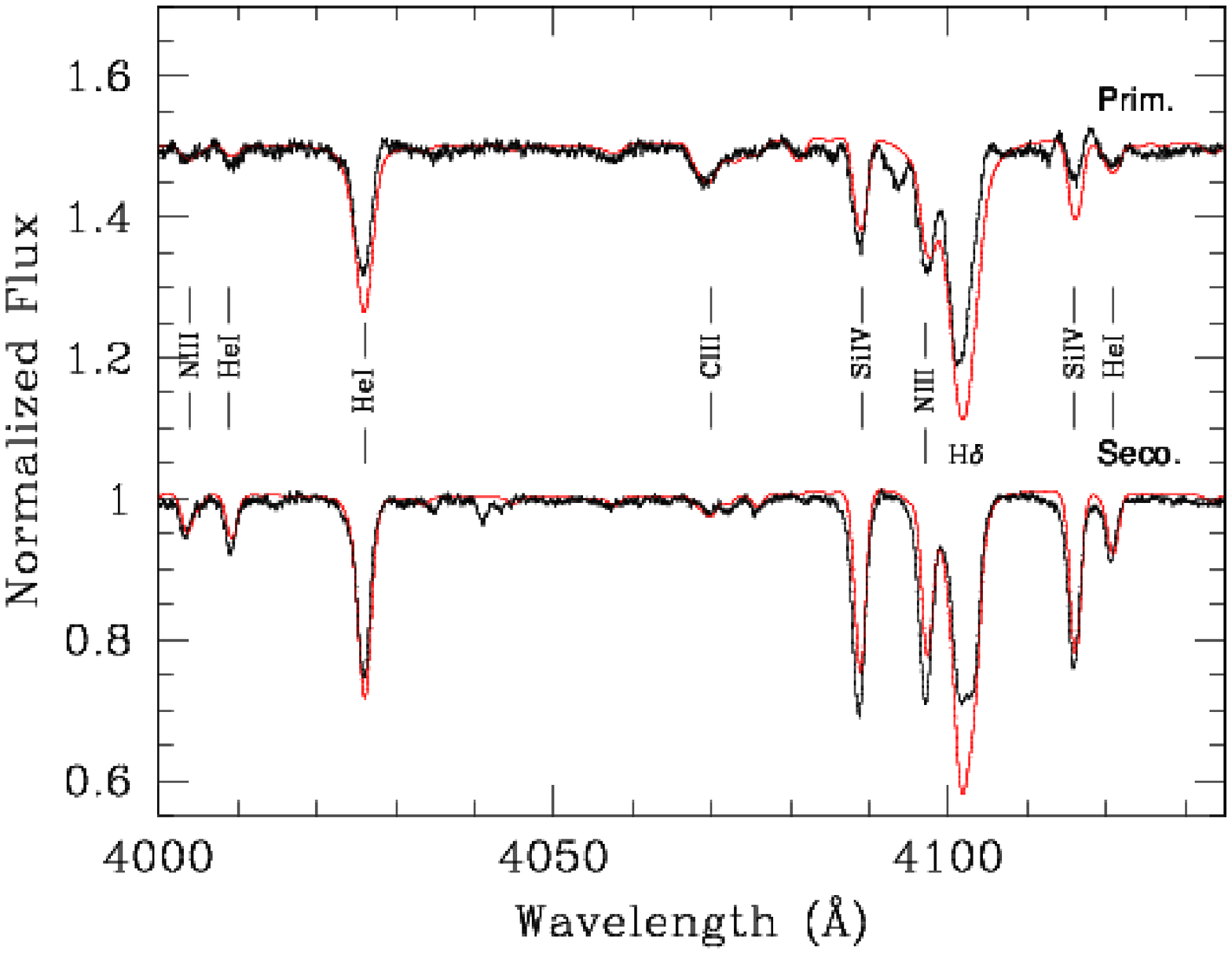}\includegraphics[scale=0.48]{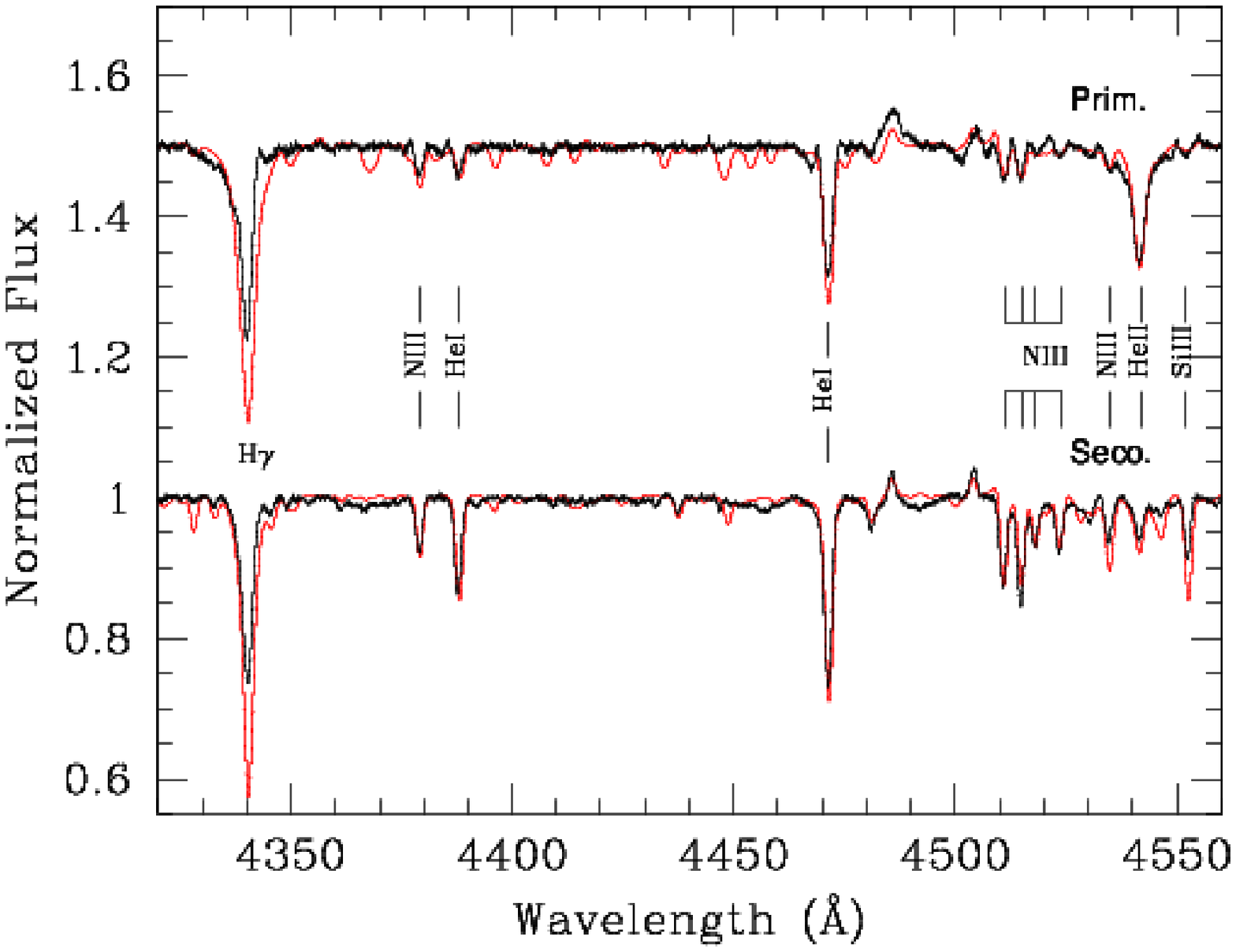}

\includegraphics[scale=0.48]{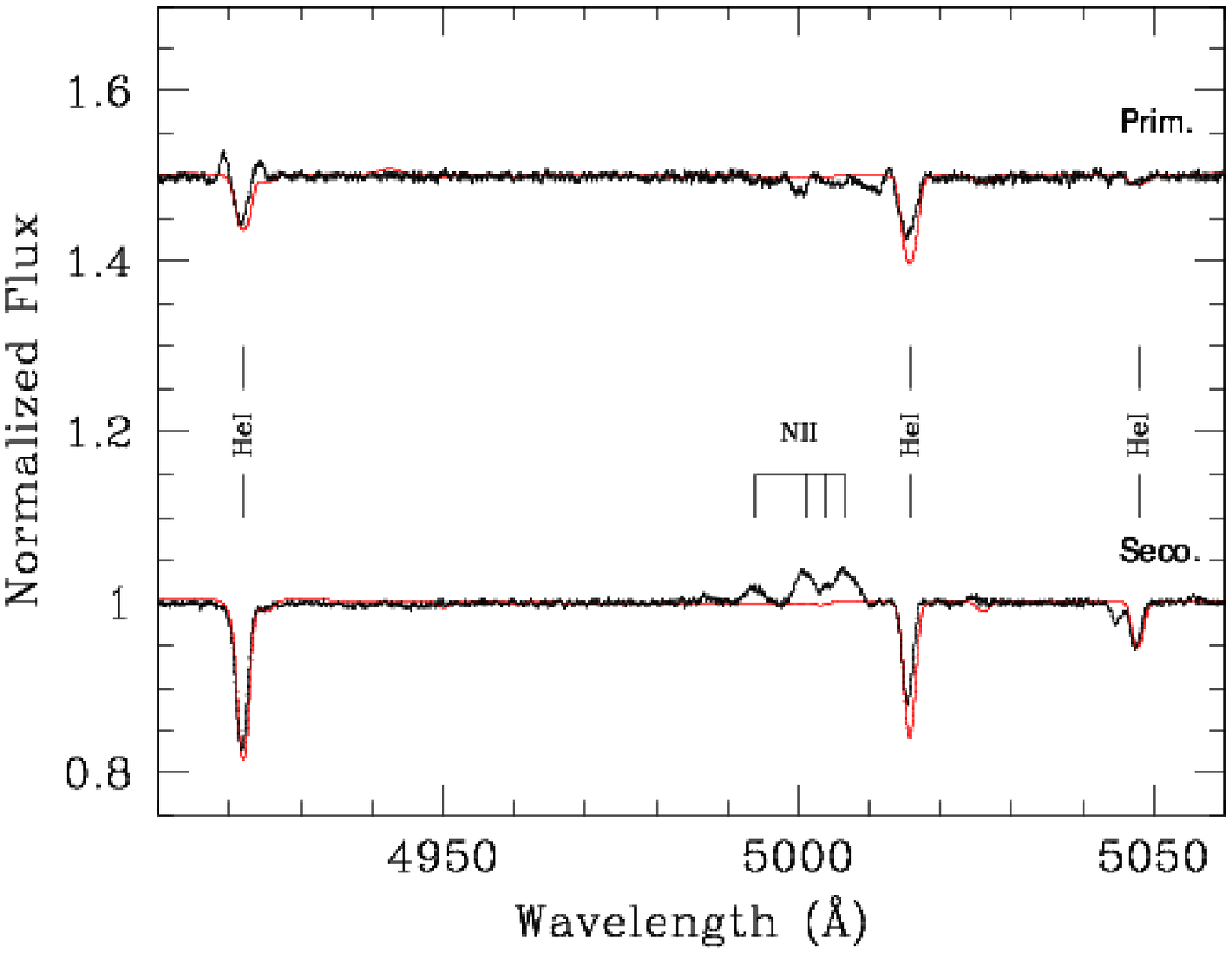}\includegraphics[scale=0.48]{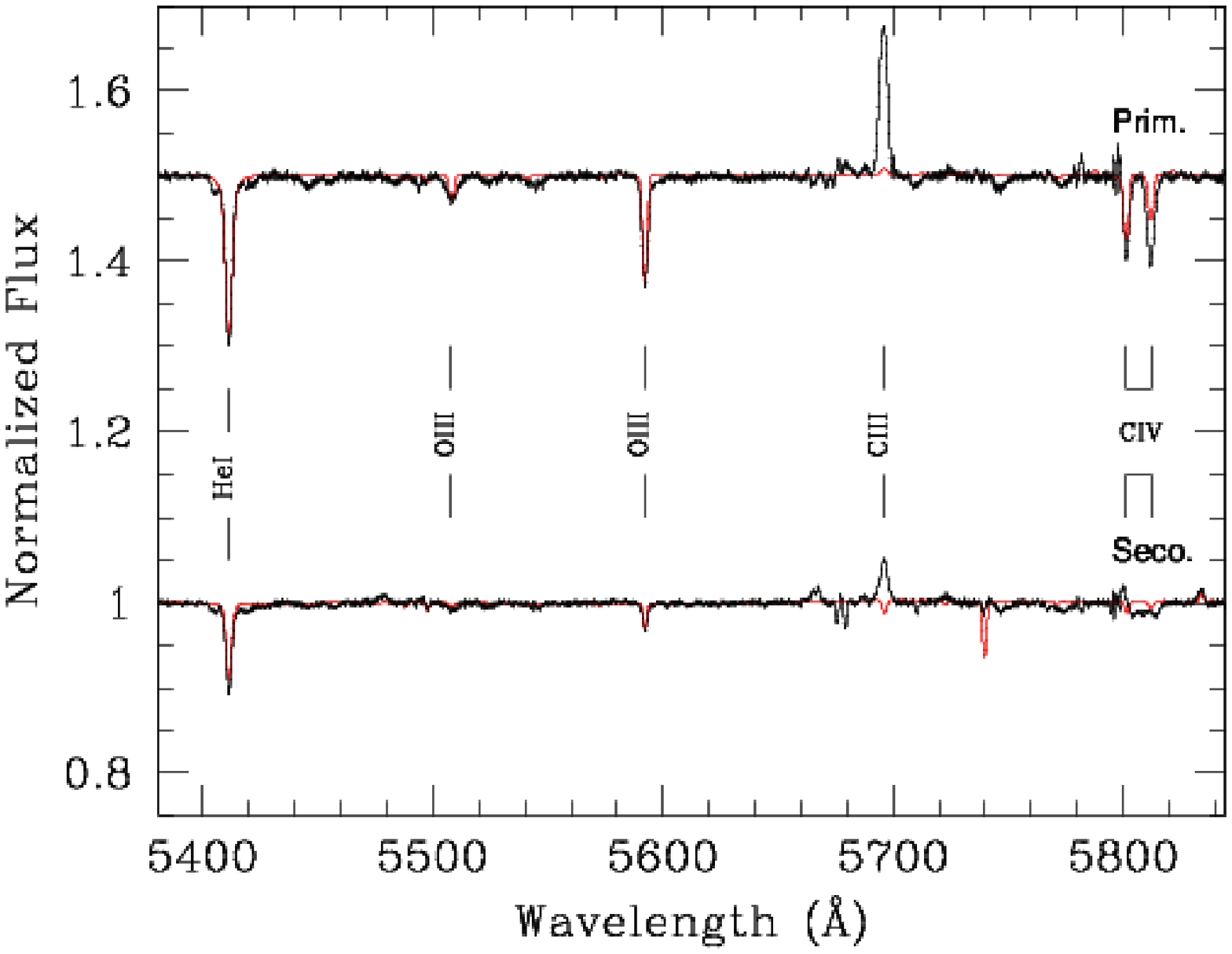}
\caption{Part of the normalized separated spectra of the primary (top, shifted
upwards by 0.5 continuum units) and secondary stars (bottom), along
with the best-fit CMFGEN model spectra (red). \label{fig1}}
\end{figure*}

Figure\,\ref{fig1} illustrates the best fit of the optical spectra
of the primary and secondary stars obtained with CMFGEN. In Table\,\ref{CMFGENparam}
we present the stellar parameters determined compared to those derived
by Rauw et al.\ (\cite{Rauw}). In general, these two works agree
well. Table\,\ref{table4} compares the chemical abundances of these
best-fit models with solar abundances taken from Asplund et al.\ (\cite{Asplund}).

Figure\,\ref{fig1} shows that the He lines and the \ion{C}{iii}
$\lambda\lambda$\,4068-70 lines are very well reproduced for both
stars. The \ion{C}{iii} $\lambda$\,5696 line cannot be considered
in the determination of the surface C abundance since its formation
process depends on a number of fine details of atomic physics and
of the modelling (Martins \& Hillier \cite{MH}). Most of the N lines
also are very well adjusted. The \ion{N}{ii} emission lines around
5000 \AA{}\ in the primary spectrum may be due to emission in the
interaction zone, as they are well present in the observed spectrum
and not represented in the modelled spectrum. The case of oxygen is
more problematic. Whilst \ion{O}{iii} $\lambda$\,5592 is well
adjusted, several other \ion{O}{iii} lines are present in the
synthetic CMFGEN spectra (e.g.\ \ion{O}{iii} $\lambda\lambda$\,
4368, 4396, 4448, 4454, and 4458), but they are barely visible in
the separated spectra. We checked that these lines are also absent
from the original observed binary spectra. Their absence is thus by
no means an artefact that is due to the treatment of our data.

\begin{table*}[!t]
\caption{Stellar parameters of the primary and secondary stars as obtained
with CMFGEN \label{CMFGENparam}}

\begin{centering}
\begin{tabular}{c|cccc}
\hline 
\multicolumn{1}{c}{} & \multicolumn{2}{c}{This study} & \multicolumn{2}{c}{Rauw et al.\ (\cite{Rauw})}\tabularnewline
\multicolumn{1}{c|}{} & Prim.  & Sec.  & Prim.  & Sec.\tabularnewline
\hline 
\hline 
$R$ (R$_{\odot}$)  & $19.3\pm2.2$  & $25.9\pm3.4$  & $24.3\pm0.7$  & $28.1\pm0.7$\tabularnewline
$M$ (M$_{\odot}$)  & $50.5\pm20.1$  & $31.9\pm9.5$  & $54.8\pm4.6$  & $33.0\pm2.8$\tabularnewline
$T_{{\rm eff}}$ ($10^{4}$\,K)  & $3.40\pm0.15$  & $2.80\pm0.15$  & $3.51\pm0.1$  & $3.05\pm0.04$\tabularnewline
log ($\frac{L}{L_{\odot}}$)  & $5.63\pm0.05$  & $5.58\pm0.04$  & $5.90\pm0.08$  & $5.78\pm0.08$ \tabularnewline
$\log{g}$ (cgs)  & $3.55\pm0.15$  & $3.05\pm0.15$  &  & \tabularnewline
$\beta$  & 1.03 {\scriptsize (f)}  & 1.08 {\scriptsize (f)}  &  & \tabularnewline
$v_{\infty}$ (km\,s$^{-1}$)  & 2450 {\scriptsize (f)}  & 2450 {\scriptsize (f)}  &  & \tabularnewline
$\dot{M}$ (M$_{\odot}$\,yr$^{-1})$  & $9.2\times10^{-7}$ {\scriptsize (f)}  & $3.3\times10^{-7}$ {\scriptsize (f)}  &  & \tabularnewline
BC  & $-3.17$  & $-2.67$  &  & \tabularnewline
\hline 
\end{tabular}
\par\end{centering}

\tablefoot{The best-fit CMFGEN model parameters are compared with
the parameters obtained by Rauw et al.\ (\cite{Rauw}) for an orbital
inclination of 21\textdegree{}. The effective temperatures from Rauw
et al.\ (\cite{Rauw}) were derived through the effective temperature
calibration of Chlebowski \& Garmany (\cite{Chle}) and permitted,
along with the determined luminosities, inferring the stellar radii.
The quoted errors correspond to 1\,$\sigma$ uncertainties. The symbol
{\small ``}{\scriptsize (f)}{\small ''} in the table corresponds
to values fixed from the literature (Howarth et al.\ \cite{Howarth};
Muijres et al.\ \cite{Muijres}). The bolometric corrections are
taken from Lanz \& Hubeny (\cite{LH}), based on our best-fit $T_{{\rm eff}}$
and $\log{g}$.} 
\end{table*}

\begin{table}
\caption{Chemical abundances of the components of HD~149\,404}

\resizebox{9cm}{!}{%
\begin{tabular}{c|ccc}
\hline 
\multicolumn{1}{c|}{} & Primary  & Secondary  & Sun\tabularnewline
\hline 
\hline 
He/H  & 0.1  & 0.1  & 0.089\tabularnewline
C/H  & $1.02_{-0.11}^{+0.10}\times10^{-4}$  & $1.89_{-0.47}^{+0.47}\ensuremath{\times10^{-5}}$  & $2.69\times10^{-4}$\tabularnewline
N/H  & $1.32_{-0.15}^{+0.20}\times10^{-4}$  & $7.15_{-1.8}^{+2.5}\times10^{-4}$  & $6.76\ensuremath{\times10^{-5}}$\tabularnewline
O/H  & $7.33_{-1.1}^{+1.1}\times10^{-4}$  & $7.85_{-1.1}^{+1.8}\ensuremath{\times10^{-5}}$  & $4.90\times10^{-4}$\tabularnewline
\hline 
\end{tabular}}

\tablefoot{Abundances are given by number as obtained with CMFGEN.
The solar abundances (Asplund et al.\ \cite{Asplund}) are quoted
in the last column. The 1$\sigma$ uncertainty on the abundances was
set to abundances corresponding to a $\chi^{2}$ of 2.0. As only one
line was considered for the C abundance of the secondary, the corresponding
calculated uncertainty was unrealistically low. A similar problem
was encountered for the O abundance of the primary star, for which
we only had two lines to work with and several absent lines over the
spectrum, as specified in the second paragraph of this section. To
circumvent these problems, we set the uncertainties for these two
abundances to values close to those calculated for the other two elements
of the corresponding star.} \label{table4} 
\end{table}

\medskip{}

We stress two interesting results from our spectral analysis. First
of all, we confirm the large overabundance of N in the secondary spectrum.
Second, we infer an asynchronous rotation of the two stars. The radii
determined with CMFGEN (Table 4), the mean projected rotational velocities
(Table 1), and the inclination estimate from Rauw et al.\ (\cite{Rauw})
yield the rotation periods of the primary and secondary stars: 3.77
$\pm$ 0.32 days and 7.46 $\pm$ 0.95 days, respectively, or a period
ratio of $0.50\pm0.11$. We note that within the uncertainties on
the inclination, the secondary rotational period is similar to the
orbital period of the system (9.81 days).We return to these points
in Sect.\,\ref{Conclusions}. Finally, we note that according to
the radii determined here, none of the stars currently fills its Roche
lobe. With the formula of Eggleton (\cite{Eggleton}), we estimate
a Roche-lobe volume filling factor of 15\% for the primary and 73\%
for the secondary.

\medskip{}

\begin{figure*}[!t]
\resizebox{8.5cm}{!}{\includegraphics{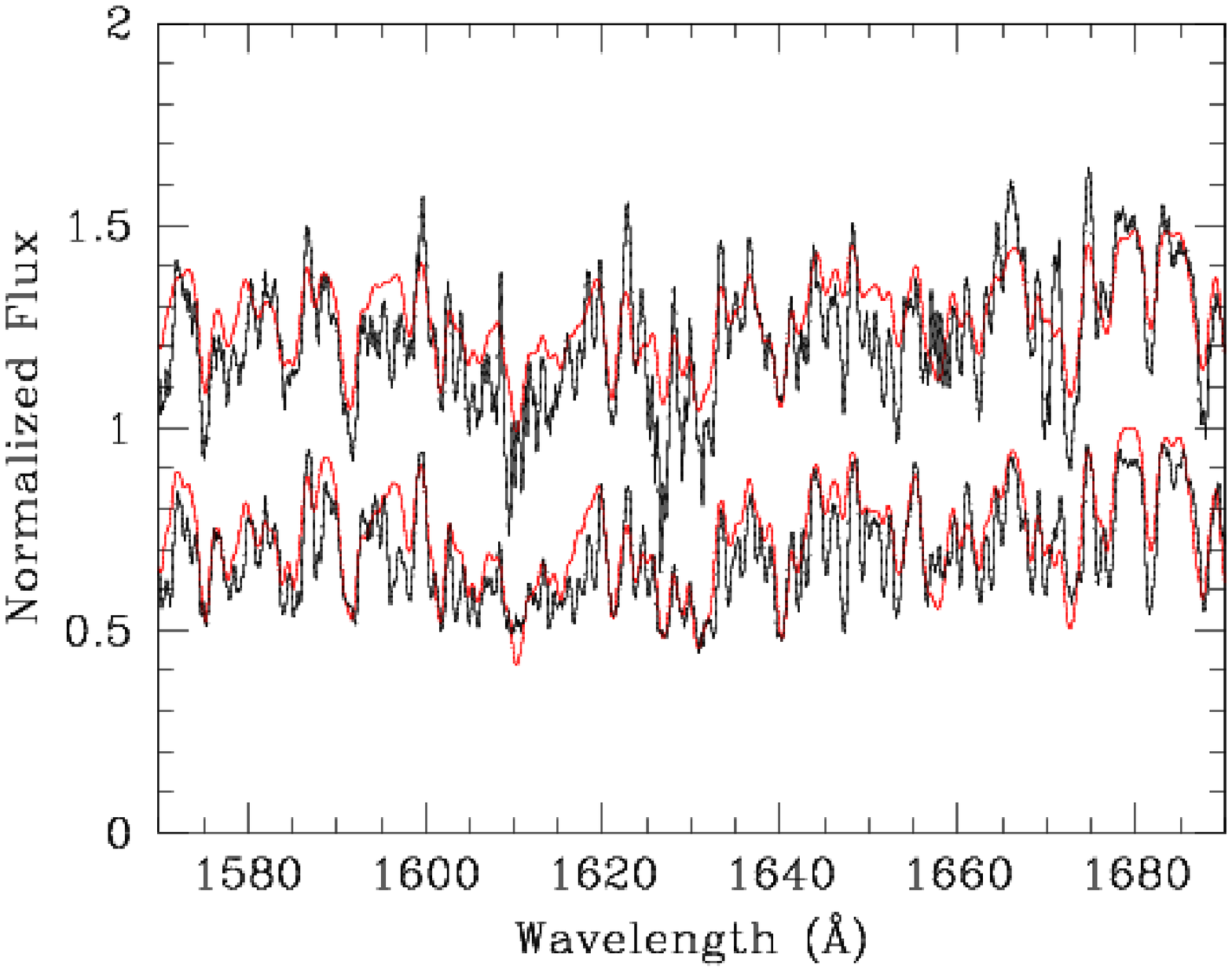}}\qquad{} \resizebox{8.5cm}{!}{\includegraphics[bb=38bp 176bp 570bp 575bp,clip]{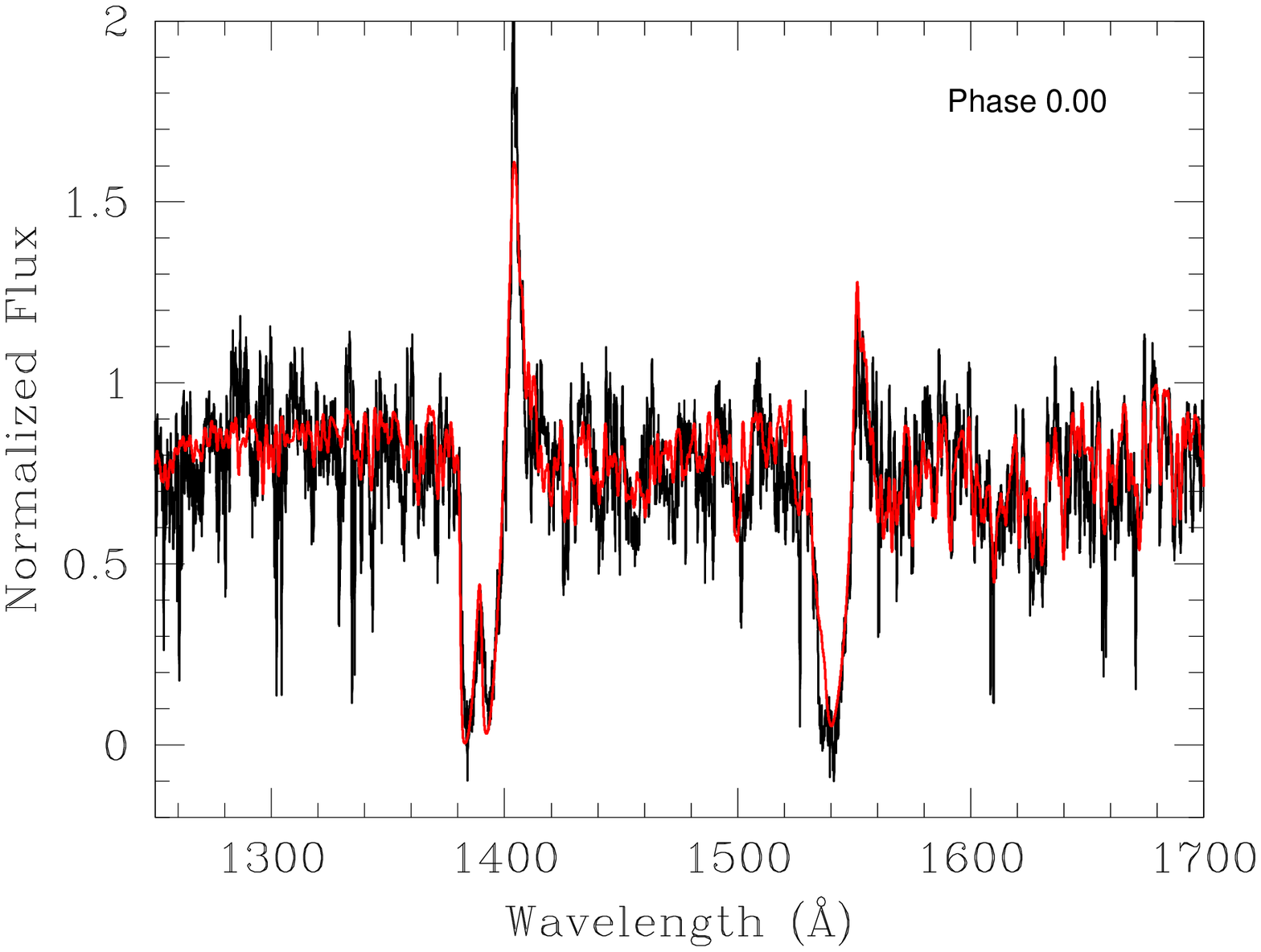}}

\resizebox{8.5cm}{!}{\includegraphics[bb=38bp 176bp 570bp 575bp,clip]{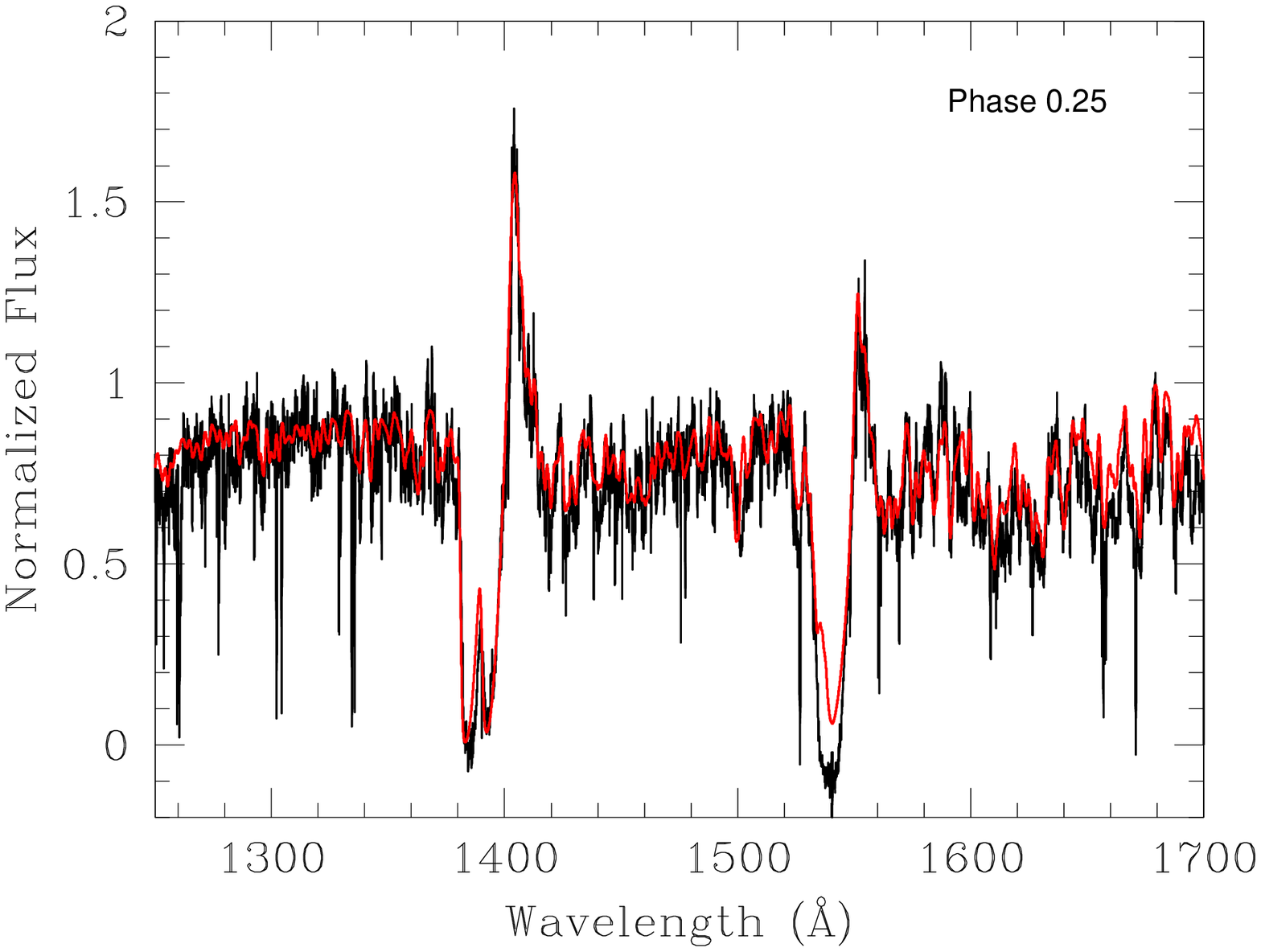}}\qquad{}
\resizebox{8.5cm}{!}{\includegraphics[bb=38bp 176bp 570bp 575bp,clip,scale=1.5]{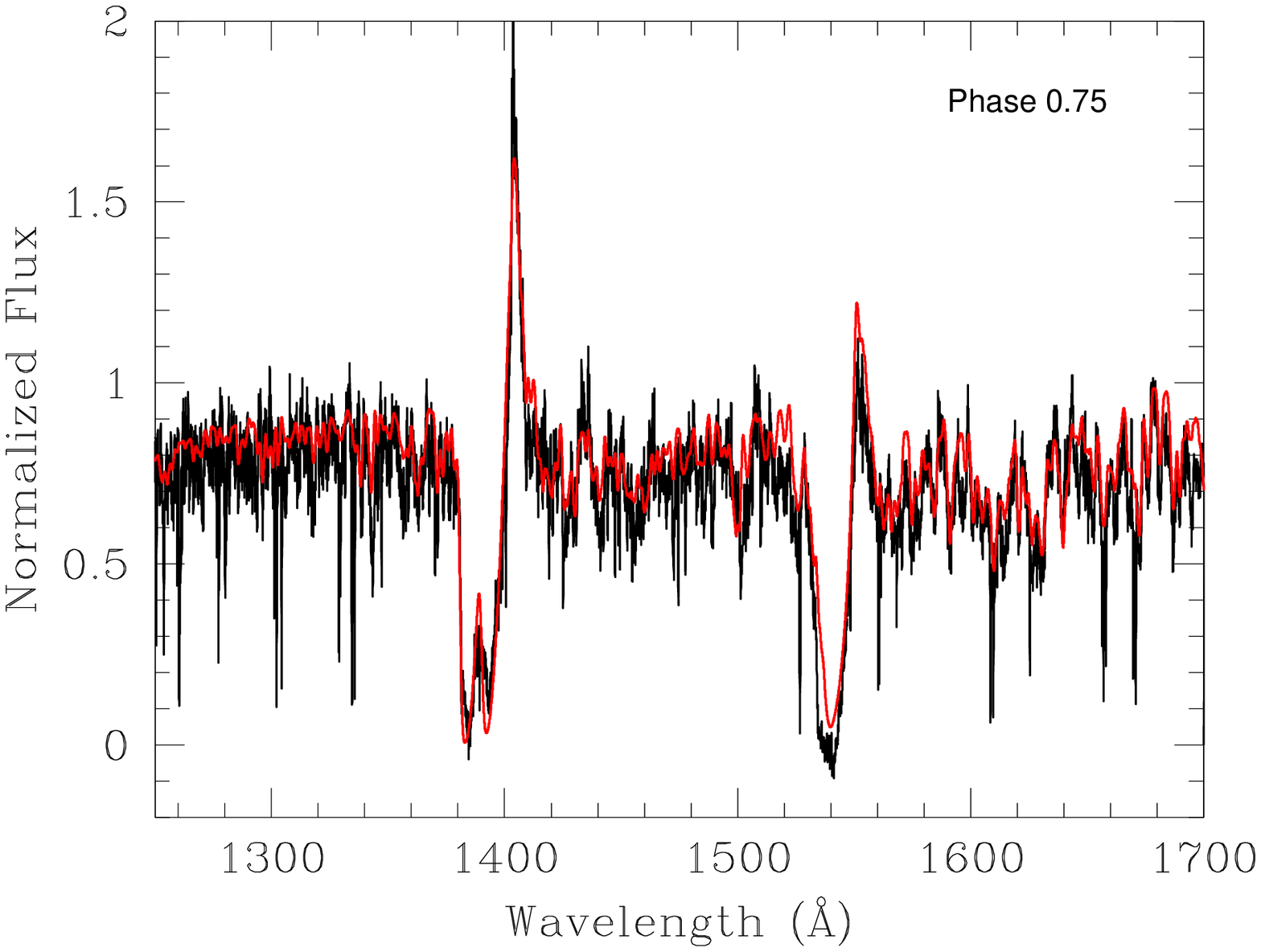}}
\caption{Upper left panel: comparison between the separated \textit{IUE} spectra
(black) of the primary and secondary stars around the \ion{He}{ii}
$\lambda$\,1640 line and the corresponding synthetic spectra (red).
The primary spectrum is shifted upwards by 0.5 for clarity. The other
three panels display a comparison between the SWP03008 (upper right),
SWP54350 (lower left) and SWP02755 (lower right) \emph{IUE} spectra
taken at phases 0.0, 0.25 and 0.75 in black and the synthetic binary
spectra obtained through combination of CMFGEN primary and secondary
spectra at given phases in red.\label{fig2}}
\end{figure*}

As pointed out above, disentangling the \textit{IUE} spectra of HD~149\,404
was more problematic, especially for those parts of the UV spectra
that feature strong and broad wind lines and/or are polluted by blends
with numerous interstellar features. The left panel of Fig.\,\ref{fig2}
illustrates the comparison of the separated spectra and the synthetic
spectra obtained with CMFGEN over a wavelength range that is relatively
free of interstellar lines and mainly hosts photospheric lines. The
agreement is clearly quite reasonable.

To check whether our CMFGEN models are also able to correctly reproduce
the wind features, we re-combined the synthetic CMFGEN UV spectra
of the primary and secondary taking into account the radial velocity
shifts at the time of the observations. We did so for several orbital
phases. The right panel of Fig.\,\ref{fig2} illustrates the result
for three selected \textit{IUE} spectra (SWP03008, JD~2\,443\,798.965,
SWP02755, JD 2\,443\,776.869, and SWP54350, JD 2\,449\,817.595),
corresponding to conjunction phase 0.00 and quadrature phases 0.75
and 0.25. The synthetic spectrum clearly agrees well with the observation,
except for a few narrow absorptions that are due to the interstellar
medium, which are not taken into account in our model. This additionally
supports our determination of the physical parameters of the system
components. Moreover, the concordance between the synthetic and observed
UV spectra suggests that the P-Cygni profiles form over parts of the
stellar winds that are not significantly affected by the wind-wind
interactions, the latter leading to a loss of spherical symmetry of
the line formation region.

The fluxed CMFGEN spectra also allow us to estimate the brightness
ratio (primary/secondary) in the UV. We find a value of 0.91%
\footnote{This result is at odds with the 0.7\,magnitude difference (i.e.\ primary/secondary
brightness ratio of 1.91) between the primary and secondary UV fluxes
inferred by Howarth et al.\ (\cite{Howarth}). This situation most
likely stems from the differences in the spectral types adopted by
Howarth et al.\ (\cite{Howarth}), who classified the system as O8.5\,I
+ O7\,III, whilst we have derived O7.5\,If + ON9.7\,I. This difference
in spectral types directly affects the strength of the cross-correlation
peaks of the spectra of the two components with the template spectrum
of $\tau$~Sco (B0.2\,IV) and thus the correction on the magnitude
difference inferred by Howarth et al. (\cite{Howarth}).%
}.

Furthermore, we found relatively little variability of the P-Cygni
profiles over the orbital period. Both observations (lack of strong
variability and little effect of wind interaction on the line morphology)
are likely due to the low orbital inclination ($i\sim21^{\circ}$)
of the system. This inclination implies that the observed spectrum
arises mostly at high stellar latitudes, where the wind-wind interaction
zone has less influence on the wind structure.

\medskip{}

\section{Discussion and conclusion\textmd{\label{Conclusions}}}

\subsection{Evolutionary status}

\begin{figure*}[htb]
\resizebox{8.5cm}{!}{\includegraphics[bb=0bp 170bp 584bp 705bp,clip]{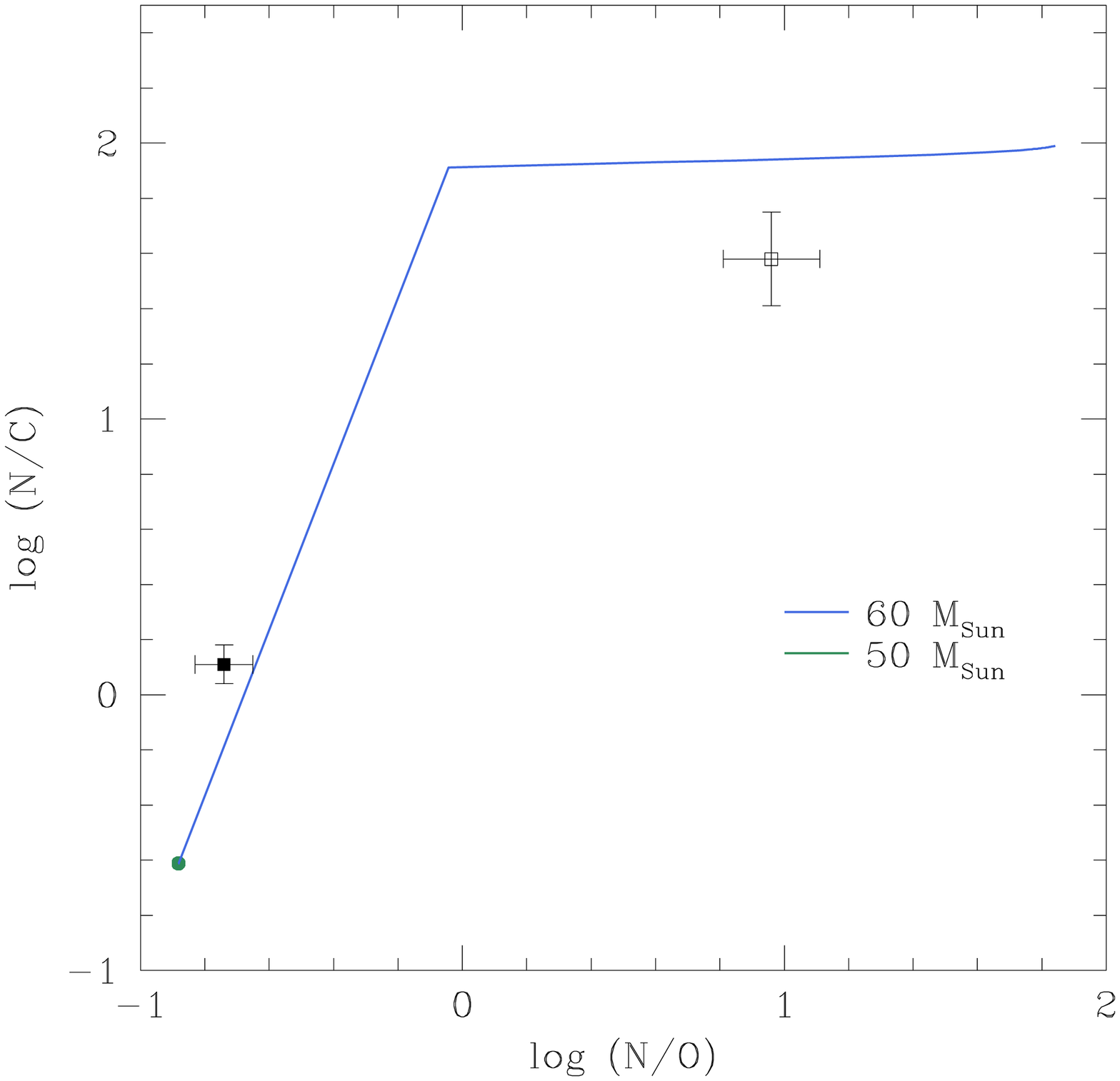}}
\resizebox{8.5cm}{!}{\includegraphics[bb=0bp 170bp 584bp 705bp,clip]{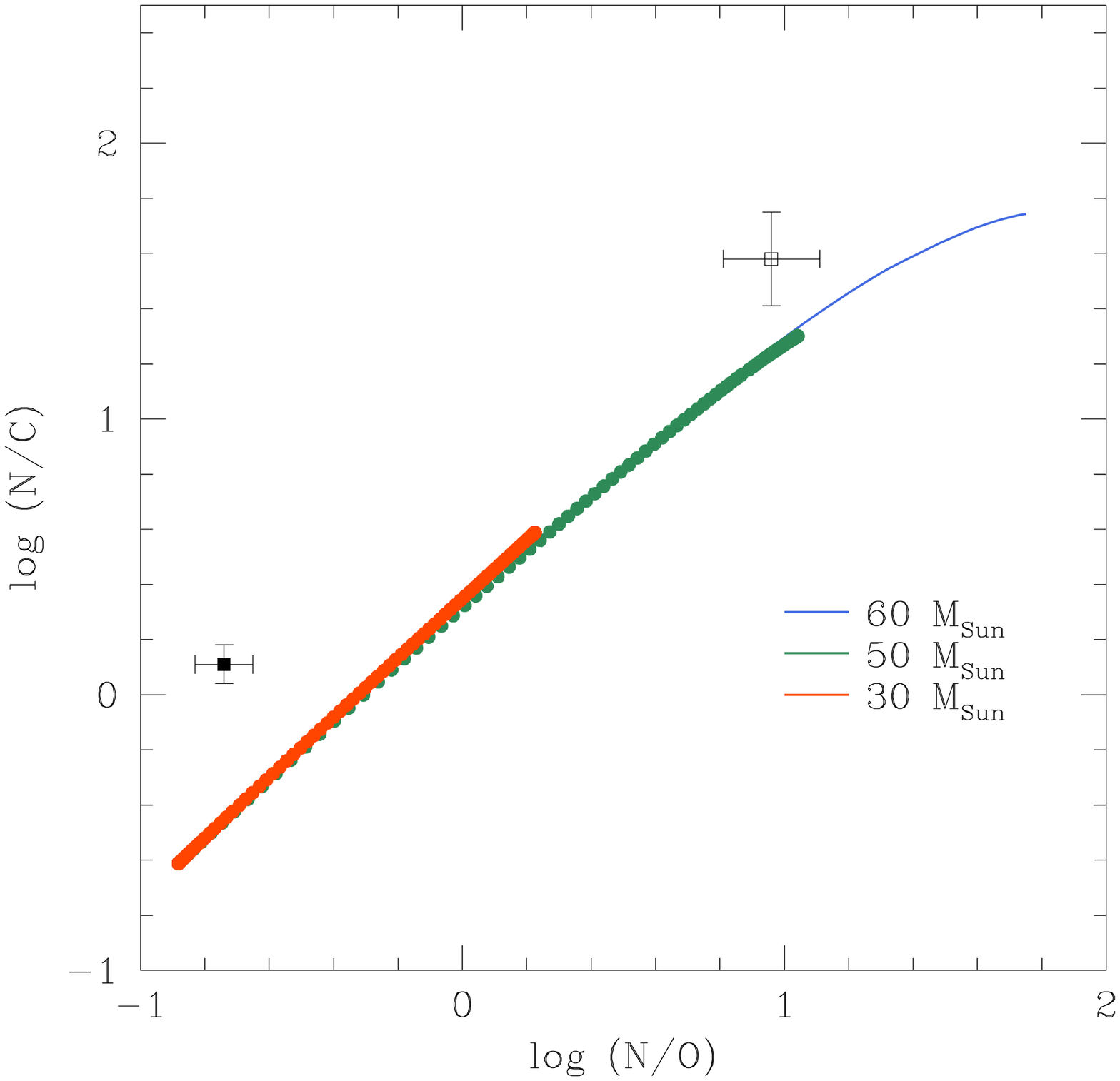}}
\caption{Comparison of the N/C and N/O ratios determined from our spectral
analyses with predictions from single-star evolutionary models of
different masses (Ekström et al.\ \cite{Ekstrom}). The primary and
secondary stars are shown by the open and filled square symbols, respectively.
The left panel illustrates the results for core hydrogen-burning phase
tracks without stellar rotation, whilst the right panel corresponds
to the same tracks for stars rotating at 0.4 $\times$ $v{}_{crit}$.\label{CNO}}
\end{figure*}

As we have shown in the previous section, the spectra of the components
of HD~149\,404 display the signatures of some enrichment by the
products of stellar nucleosynthesis processes. These properties indicate
a previous mass-exchange episode. Figure\,\ref{CNO} compares our
inferred N/C and N/O ratios with the predictions for the evolution
of single massive stars (Ekström et al.\ \cite{Ekstrom}) either
without rotation (left panel) or with a rotational velocity of $0.4\times v_{crit}$
(right panel). This figure clearly shows that for the spectroscopic
masses of the primary and secondary stars these models cannot account
for the observed CNO abundance patterns. For instance, in the case
of models without rotation (left-hand panel of Fig.\,\ref{CNO}),
no chemical enrichment is expected during the main-sequence evolution
for stars with mass $<60$\,M$_{\odot}$. Only for the M$\,>60\, M_{\odot}$
model is the loss of material through stellar wind sufficient to reveal
the products of the CNO cycle at the surface before the end of the
main-sequence evolution. Adopting instead the rotating models (right-hand
panel of Fig.\,\ref{CNO}), the enrichment of the secondary can only
be explained by single-star evolution if we assume at least a star
with 60\,M$_{\odot}$ , whilst the spectroscopic mass is about a
factor two lower. When the surface abundances of a single star of
60\,M$_{\odot}$ initial mass reach the values observed for the secondary
of HD~149\,404, the current mass of the star is still about 50\,M$_{\odot}$,
that is, it is\ much higher than the mass determined for the secondary
star.

The observed CNO abundance pattern is qualitatively much easier to
interpret if the present-day secondary star was formerly the more
massive component of the binary and transferred mass and angular momentum
to the present-day primary in an RLOF episode. During this process,
the outer envelope of the secondary was removed, revealing layers
of material that were previously inside the convective hydrogen-burning
core (Vanbeveren \cite{Vanbeveren1,Vanbeveren}).

We start by comparing our results with theoretical predictions from
the literature. The expectation is that the more mass has been lost
by the donor, the deeper the layers that are revealed at the surface
and thus the higher the N/C ratio. The predicted CNO abundances for
the mass loser after a case B mass-transfer episode are $\left[{\rm N/C}\right]=100\,\left[{\rm N/C}\right]_{\odot}$
and $\left[{\rm N/O}\right]\geq5\,\left[{\rm N/O}\right]_{\odot}$
(Vanbeveren \cite{Vanbeveren1}). This is indeed close to what we
observe in our analysis. Moreover, the RLOF scenario is re-enforced
by the values of the primary star abundances. Classical radiative
equilibrium evolutionary models predict normal H and He abundances
for the mass gainer of an RLOF episode, but an N abundance of two
to three times solar (Vanbeveren \& de Loore \cite{VdL}), which agrees
with our results.

There are some caveats here, however. First, the results of Vanbeveren
(\cite{Vanbeveren1}) apply to case B mass transfer, whereas HD~149\,404
is more likely to have undergone a case A mass exchange (see below).
Second, when the outer hydrogen-rich layers are removed, an enrichment
of the stellar surface of the mass donor in helium is expected. Our
spectral modelling was made with the He abundance set to solar, however,
and there is no hint for our models requiring a He enhancement in
any of the stars. Although the predicted relative change of helium
surface abundance is usually much smaller than for N, this result
is nonetheless surprising, especially in view of the large N overabundance
of the secondary star.

To illustrate the helium enrichment that might be expected, we consider
once more the single-star evolutionary tracks with rotation of Ekström
et al.\ (\cite{Ekstrom}). In these models, a nitrogen surface abundance
similar to the one observed for the secondary is always accompanied
by an increase of the surface helium abundance to values higher than
$y=0.35$ by mass. The fact that the spectrum of the secondary of
HD~149\,404 can be fit with a solar helium abundance ($y=0.1$)
suggests that the internal structure of the secondary before the onset
of mass transfer must have been different from that expected from
single-star evolutionary models. Moreover, the RLOF in HD~149\,404
must have come to a (temporary) stop before the full hydrogen-rich
envelope of the mass donor was removed. The latter conclusion is also
supported by our finding that both stars are currently well inside
their Roche lobes.

Given the present-day spectroscopic masses, the secondary must have
lost at least 9.3\,M$_{\odot}$ assuming a fully conservative mass
transfer. Conservative RLOF will increase the period of the binary.
If we assume a fully conservative mass transfer and apply the formula
of Vanbeveren et al.\ (\cite{VDLVR}), we find that the orbital period
would have increased by 17\% at most, the shortest possible initial
orbital period being 8.4\,days for a mass exchange of 9.3\,M$_{\odot}$.
However, Petrovic et al.\ (\cite{Petrovic}) argued that in fast
case A RLOF, only about 15 -- 20\% of the mass shed by the mass donor
is indeed accreted by its companion. If the mass-transfer efficiency
is that low, then the secondary must have lost a much larger amount
of mass. Such a non-conservative mass transfer can result from the
transfer of angular momentum. In fact, the mass gainer can be spun
up to critical rotation and will then repel additional mass (and angular
momentum), leading to a non-conservative mass transfer (e.g.\ Langer
et al.\ \cite{Langer}, Vanbeveren \cite{Vanbeveren}, Langer \cite{Langer1}).
In the previous section, we noted that the components of the system
are apparently not in synchronous rotation. This asynchronicity therefore
also suggests the existence of a past mass and angular momentum transfer
episode, with a clear spin up of the primary star. %from the secondary star to the primary one and could hint at an alteration of the spin due to the transfer of angular momentum. 

%Note that this potential initial orbital period is similar to the secondary rotational period (7.46 days).

\begin{figure*}[htb]
\resizebox{8.5cm}{!}{\includegraphics[bb=40bp 175bp 580bp 711bp,clip]{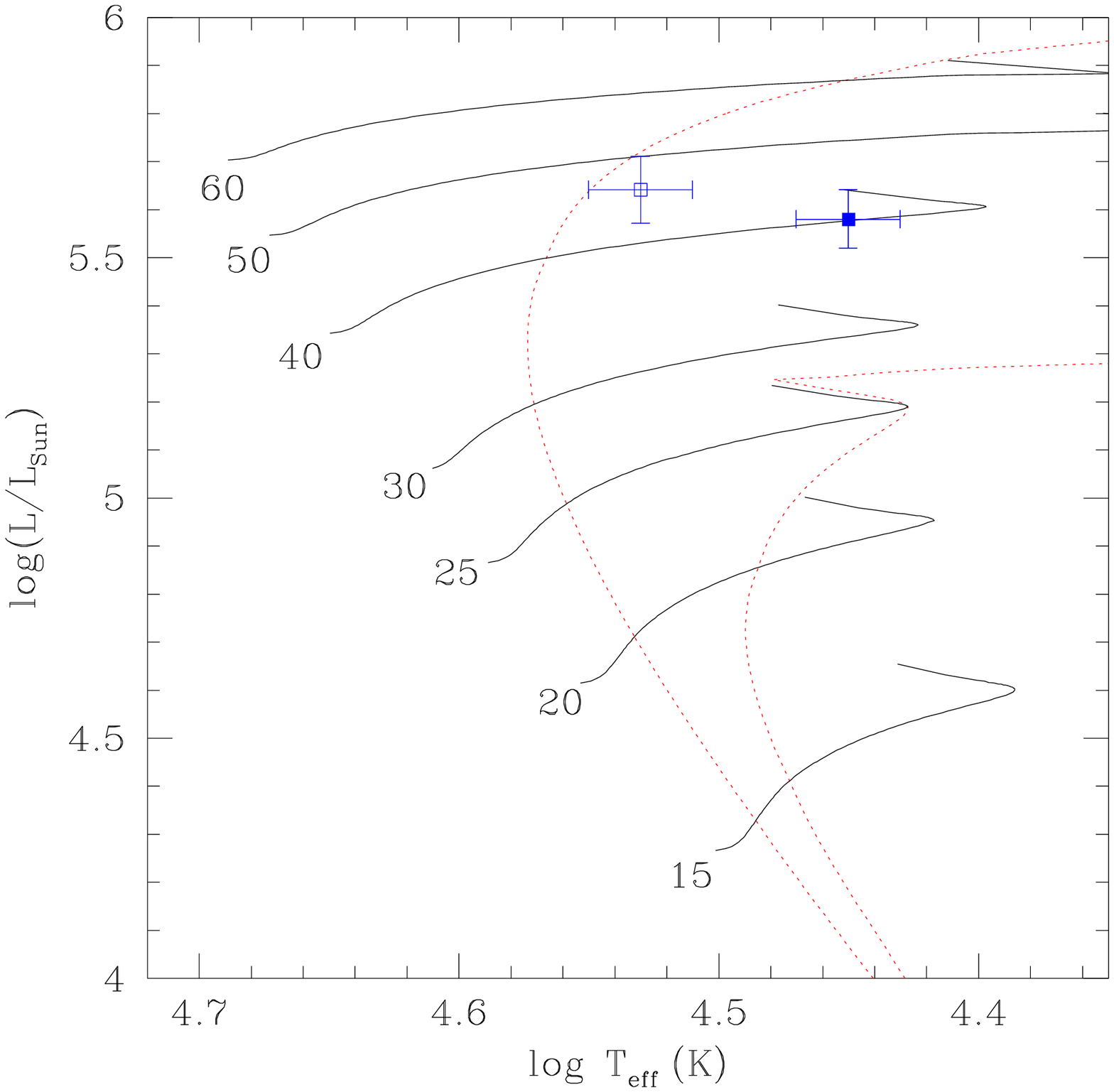}}\enskip{}
\resizebox{8.5cm}{!}{\includegraphics[bb=40bp 175bp 580bp 711bp,clip]{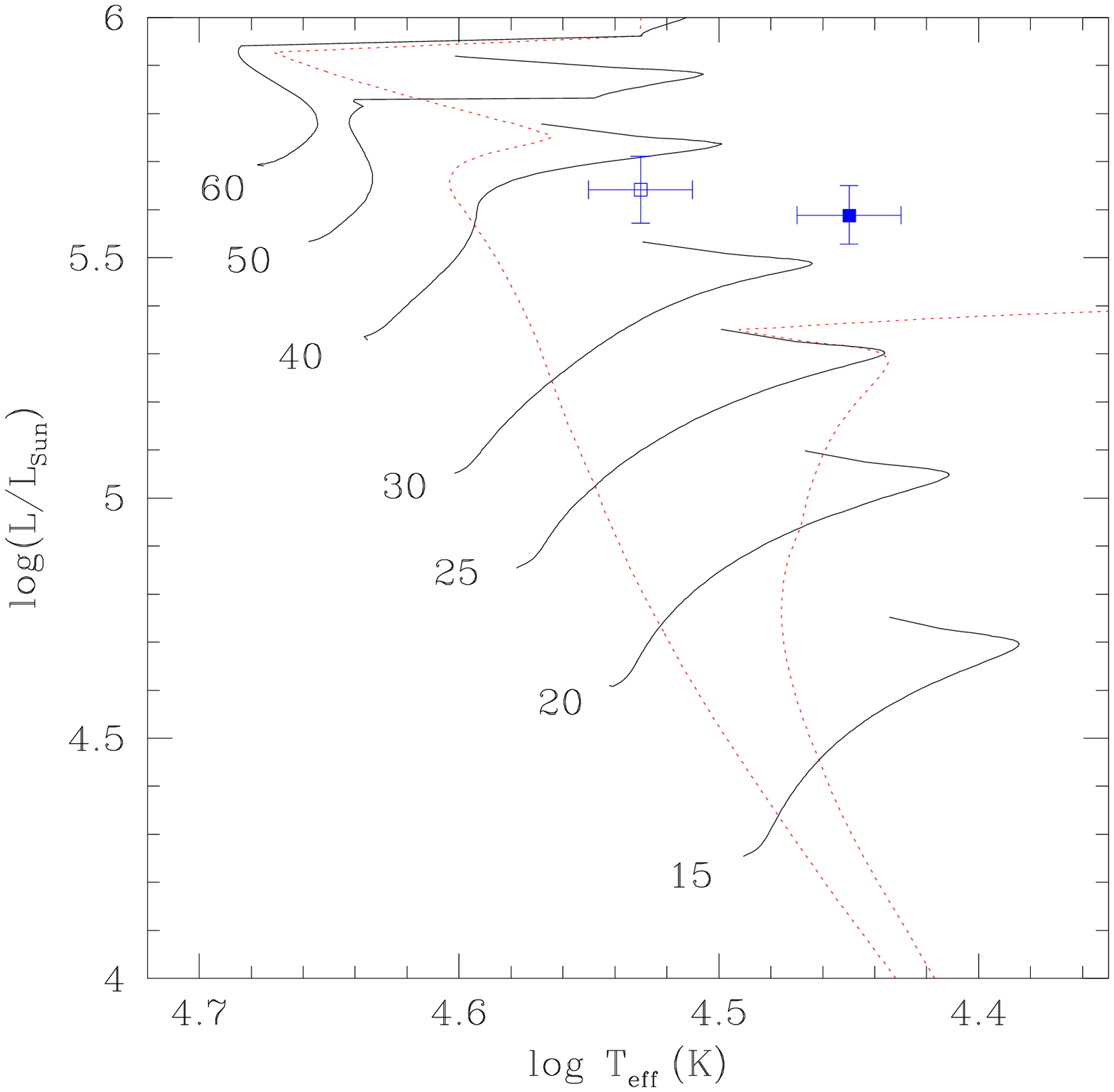}}

\medskip{}

\resizebox{8.5cm}{!}{\includegraphics[bb=38bp 178bp 580bp 711bp,clip]{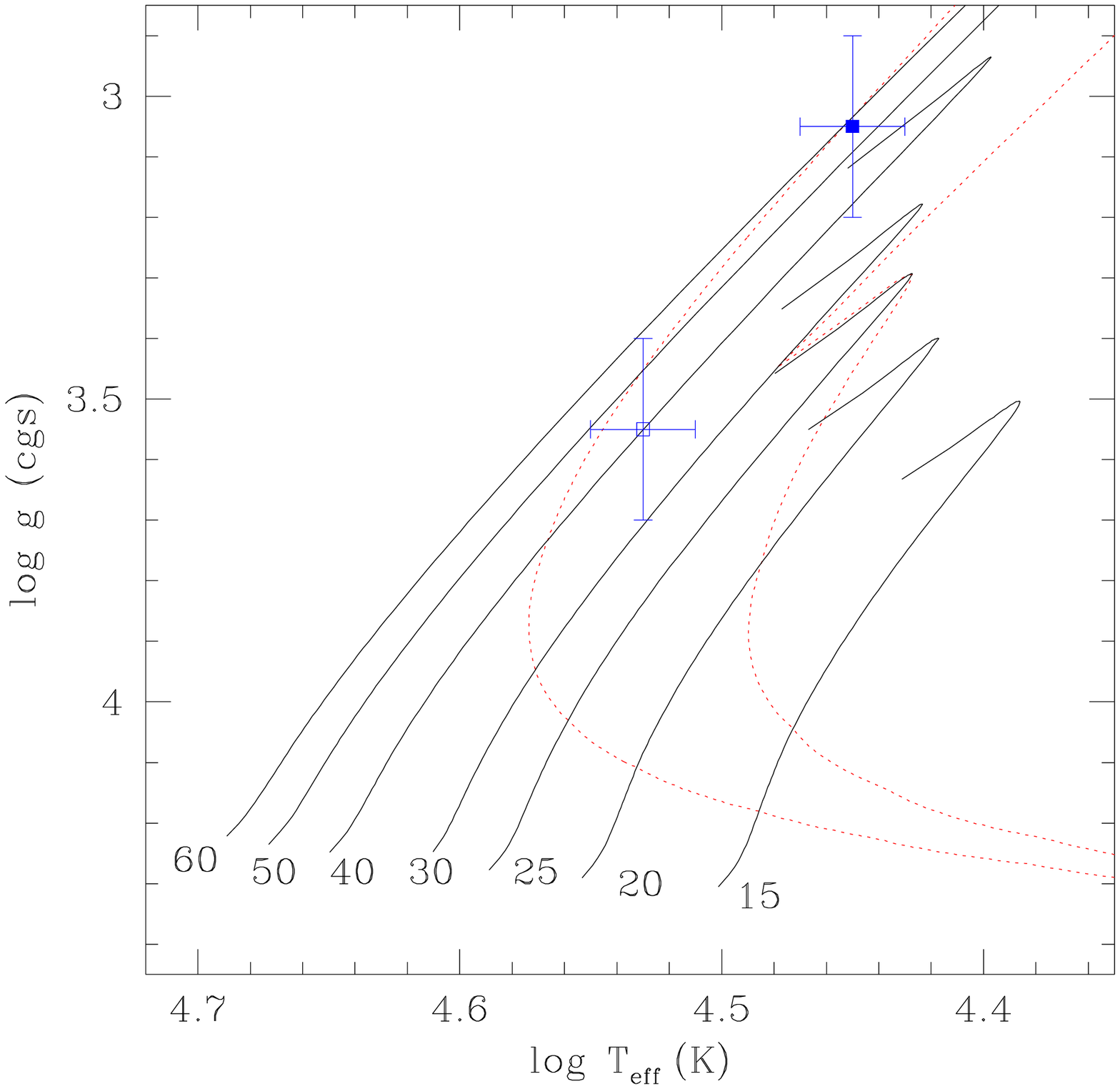}}\enskip{}
\resizebox{8.5cm}{!}{\includegraphics[bb=38bp 178bp 580bp 711bp,clip]{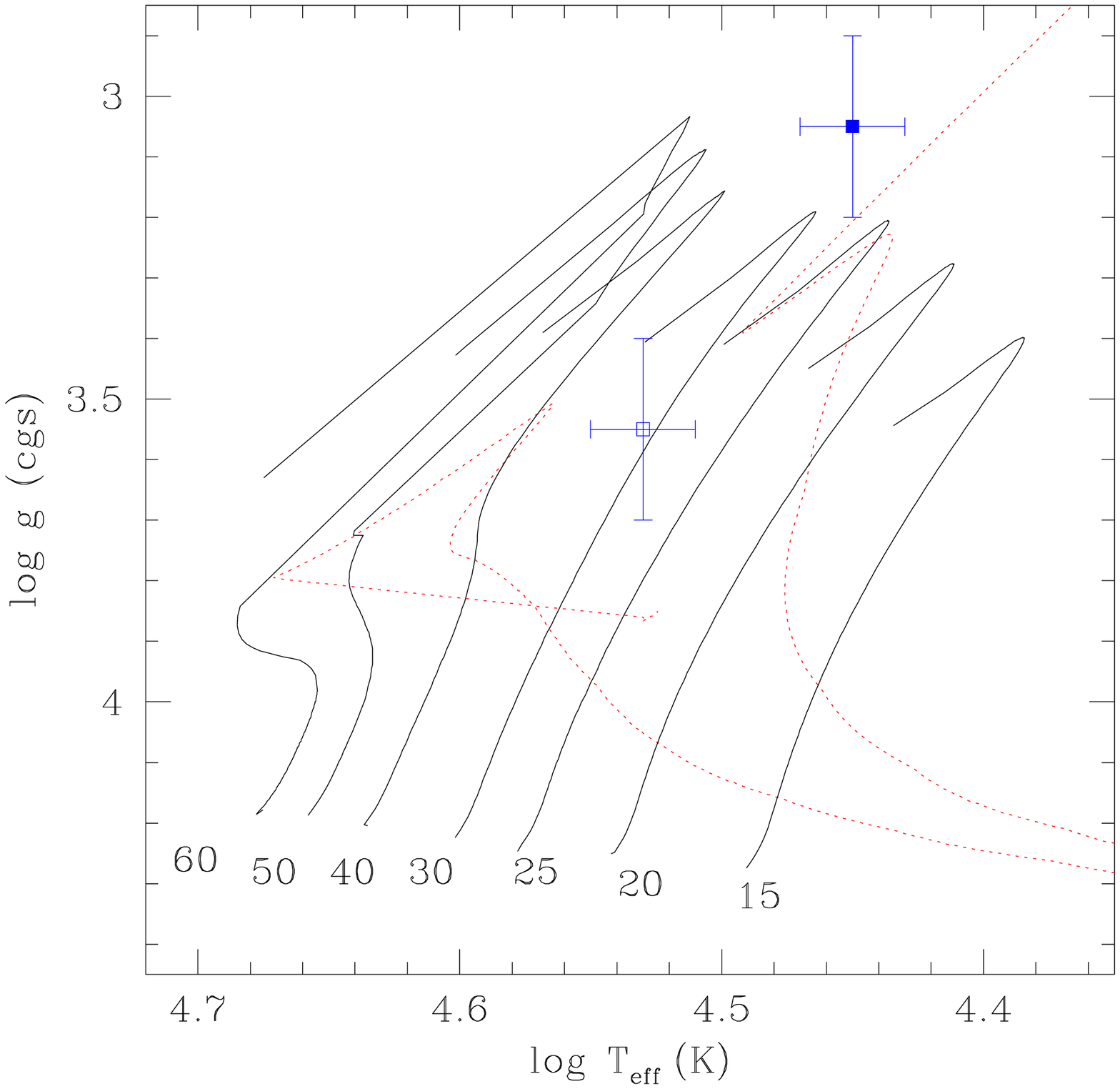}}

\caption{Position of the primary (open square) and secondary (filled square)
stars in the Hertzsprung-Russell diagram (upper panels) and the log\,g-log\,$T_{{\rm eff}}$
diagram (lower panels) along with evolutionary tracks for single massive
stars at solar metallicity during the core-hydrogen burning phase
(Ekström et al.\ \cite{Ekstrom}). In the left panels, the evolutionary
tracks correspond to non-rotating stars, whilst the right panels yield
the results for stars rotating at 0.4 $\times$ $v{}_{crit}$. The
dotted red lines correspond to isochrones of 3.2 and 6.3\,million
years for the left panels and of 4.0 and 8.0\,million years for the
right panels.\textbf{ \label{HRD}}}
\end{figure*}

In Fig.\,\ref{HRD} we present the positions of the primary and secondary
stars in the Hertzsprung-Russell diagram (HRD) (upper panels) and
in the log\,g-log\,$T_{{\rm eff}}$ diagram (lower panels), along
with the evolutionary tracks from Ekström et al.\ (\cite{Ekstrom})
for the core-hydrogen burning phase of single stars at solar metallicity,
both for non-rotating (left panels) and rotating (right panels) stars.
This figure again illustrates the failure of single-star evolutionary
tracks to account for the properties of the components of HD~149\,404.
If we consider the models without rotational mixing, the position
of the primary star in the HRD is relatively close to the track corresponding
to its present-day spectroscopic mass. However, the secondary appears
too luminous for its present-day mass compared to these evolutionary
tracks. If we compare the stars to the 40\% critical rotation case,
we find that the primary's luminosity is below that expected for its
present-day spectroscopic mass. Moreover, the secondary lies outside
the range covered by tracks in the core-hydrogen burning phase. If
we consider the non-rotating case, we see that in the log\,g-log\,$T_{{\rm eff}}$
diagram the primary star has a lower effective temperature and a higher
surface gravity than expected for a 50 M$_{\odot}$ star, whilst the
secondary star displays parameters corresponding to stars twice as
massive as its present-day spectroscopic mass. For stars rotating
at 40\% of their critical velocity, we find that both stars would
have initial masses of about 30 M$_{\odot}$. The primary dynamical
mass is about a factor 2 higher, while for the secondary the agreement
is good. Finally, in all cases (with and without rotation and for
both diagrams), the two components of HD~149\,404 are not located
on the same isochrone, although they would be expected to have the
same age%
\footnote{The age of the Ara\,OB1 association of which HD~149\,404 is a member
was estimated in the literature as $\sim3$ Myr (Wolk et al.\ \cite{Wolk}).%
}.

\medskip{}

In a mass-transfer episode, the location of both stars in the HRD
after the RLOF strongly depends on the details of the process. For
instance, spinning-up the mass gainer can lead to almost complete
mixing of the latter. In this situation, models predict that He would
be enhanced in the atmosphere of the mass gainer and that the latter
would become bluer and overluminous for its mass (Vanbeveren \& de
Loore \cite{VdL}). If there is no strong spin-up and the accretion
occurs rather slowly and is accompanied by a fast diffusion process,
then the location of the gainer in the HRD should be very close to
that of a normal star of the same mass and chemical composition (Vanbeveren
et al.\ \cite{VDLVR}). The star is rejuvenated, however: it lies
on an isochrone that corresponds to a younger age than its actual
age. The location of the primary in the HRD suggests that in the present
case the second scenario is likely to apply with a very moderate rejuvenation,
if any, of the primary. This conclusion is supported by the lack of
a strong He enrichment, but it is at odds with the highly non-synchronous
rotation of the primary reported above.

\medskip{}

\subsection{Comparison with other post-RLOF systems}

In this section, we briefly compare the properties of HD~149\,404
with those of two other O-type binaries, Plaskett's Star and LZ\,Cephei,
which are probably in a post-case A RLOF evolutionary stage and have
been analysed in a way similar to the present work (Linder et al.\ \cite{Linder},
Mahy et al.\ \cite{Mahy1}).\\

Plaskett's Star (HD~47\,129) is an O8\,III/I + O7.5\,III binary
with an orbital period of 14.4\,days, slightly longer than that of
HD~149\,404, and a mass ratio near unity. Two observational facts
suggest that the initially more massive primary has undergone a case
A RLOF episode in the past (Linder et al.\ \cite{Linder}). First,
the projected rotational velocity of the secondary star is remarkably
high (mean value of $272.5\pm35.0$\,km\,s$^{-1}$) compared to
that of the primary (mean value of $65.7\pm6.3$\,km\,s$^{-1}$).
This strongly asynchronous rotation likely stems from a spin-up of
the secondary star through a recent transfer of angular momentum from
the primary star. Second, the primary star displays nitrogen overabundance
(N/N$_{\odot}=16.6\pm5.0$) and carbon depletion (C/C$_{\odot}=0.2\pm0.1$),
as is typical of an advanced evolutionary stage, whilst no such nitrogen
enrichment is seen in the secondary spectrum%
\footnote{The apparent N underabundance of the secondary inferred by Linder
et al.\ (\cite{Linder}) could be an artefact that is due to the
relatively shallow nitrogen lines of the secondary being washed out
by its fast rotation and the difficulties of properly normalizing
echelle spectra with such broad and shallow features (see also Fig.\,1
of Palate \& Rauw \cite{PR}).%
}. Given the present-day configuration of HD~47\,129, it seems that
the transfer of angular momentum and the resulting spin-up of the
mass gainer (the secondary in this case) might have prevented accretion
of a larger portion of the primary's mass.\\

LZ\,Cephei (HD~209\,481) is an O9\,III + ON9.7\,V binary with
a short orbital period of 3.07\,days (Kameswara Rao \cite{Rao})
and a mass ratio (primary/secondary) of 2.53 (Mahy et al.\ \cite{Mahy1}).
Both components of LZ\,Cep almost fill their Roche lobe (Mahy et
al.\ \cite{Mahy1}, Palate et al.\ \cite{PRM}). Mahy et al.\ (\cite{Mahy1})
found that the secondary displays a strong nitrogen enrichment (18
times solar), carbon and oxygen depletion (0.1 times solar), and a
high helium content (4 times solar), suggesting a very advanced state
of evolution. On the other hand, the primary displays only slightly
altered surface abundances consistent with single-star evolution.
This suggests that the mass transfer was strongly non-conservative.
Mahy et al.\ (\cite{Mahy1}) noted a slight asynchronicity of the
rotation periods of the order of 10-20 \%, but concluded that the
system is probably on the verge of achieving synchronization. Comparison
with single-star evolutionary tracks indicates a similar age discrepancy
as we have found here for HD~149\,404. The evolutionary masses ($25.3_{-4.0}^{+6.2}$
and $18.0_{-2.5}^{+2.4}$\,M$_{\odot}$, for the primary and secondary
respectively) are significantly higher than their dynamical masses
($16.9\pm1.0$ and $6.7\pm1.0$ M$_{\odot}$ respectively, Mahy et
al.\ \cite{Mahy1}, Palate et al.\ \cite{PRM}).\\

Although our sample is currently too limited to draw firm conclusions,
the above comparison reveals several interesting results that are
worth studying in more detail with a larger sample. First, we note
that in all three systems, the components appear either over- or under-luminous
when compared with single-star evolutionary tracks corresponding to
their present-day masses. During its initial phases, case A mass transfer
occurs on a thermal timescale leading to a very rapid mass exchange.
However, the mass transfer subsequently slows down and occurs on a
nuclear timescale instead. By the time the mass transfer ends, both
stars would therefore be expected to be in thermal equilibrium again.
Since all three systems under consideration are in a detached configuration,
the stars should thus have settled onto their new post-RLOF evolutionary
tracks. The discrepancy between the observed luminosities and those
predicted by the single-star evolutionary tracks most likely stems
from several effects, such as observational uncertainties, altered
chemical compositions (and hence opacities of the stellar material),
and a rotational velocity history that differs from that assumed in
the single-star models.

Our comparison shows that the degree of asynchronicity in the three
post-RLOF systems increases with orbital period (from LZ\,Cep, over
HD~149\,404 to HD~47\,129). This is most probably due to the higher
efficiency of the tidal forcing in shorter period systems, or it might
be seen as evidence that momentum transfer has played a larger role
in those systems that have a wider present-day orbital separation.
This is expected since binary evolution models predict spin-up to
be less efficient in closer systems because in such systems the accretion
stream directly affects the mass gainer without forming an accretion
disk (e.g.\ Langer \cite{Langer1}). The questions of synchronization
and circularization in close binary systems with early-type stars,
featuring radiative envelopes and convective cores, have been addressed
by Zahn (\cite{Zahn1}) and Tassoul (\cite{Tassoul}). Zahn (\cite{Zahn1})
considered radiative damping of the dynamical tide in the outer layers
of the stellar envelopes as the main mechanism for synchronization.
The corresponding synchronization timescales strongly depend upon
the tidal torque constant $E_{2}$ , which varies by several orders
of magnitude during the main-sequence lifetime of the star (Zahn \cite{Zahn2}).
Siess et al.\ (\cite{Siess}) provided a parametrization of this
parameter as a function of stellar mass and relative age of the star.
Assuming that their formalism holds for the masses of the stars considered
here and adopting an age of half the main-sequence age of the stars,
we estimated the values of $E_{2}$. We then evaluated the synchronization
times following Eq. 4.28 of Zahn (\cite{Zahn1}) by means of moments
of inertia tabulated by Claret \& Giménez (\cite{CG}). The results
are $5.6\times10^{5}$ and $1.4\times10^{4}$\,years for the primary
and secondary stars of HD~149\,404, respectively, $4.4\times10^{6}$
and $27.6\times10^{6}$\,years for the primary and secondary of Plaskett's
star, and $2.8\times10^{4}$ and $2.5\times10^{4}$\,years for the
primary and secondary components of LZ~Cephei. Tassoul (\cite{Tassoul}),
on the other hand, proposed that the most effective mechanism for
synchronization is provided by large-scale hydrodynamic motions in
the interiors of the tidally distorted binary components. The resulting
timescales are considerably shorter than those estimated following
the formalism of Zahn (\cite{Zahn1}). We estimate synchronization
timescales of $2.1\times10^{4}$ and $3.2\times10^{3}$\,years for
HD~149\,404, $7.2\times10^{4}$ and $2.0\times10^{5}$\,years for
Plaskett's star, and $1.6\times10^{3}$ and $6\times10^{2}$\,years
for LZ~Cep. Regardless of which of the two formalisms is most appropriate
to describe the evolution of rotation in the systems considered here,
we find that, except perhaps for Plaskett's Star, which has a longer
synchronization timescale, the RLOF phase probably must have ended
less than a few $10^{4}$\,years ago. %Our finding is also in line with the results of de Mink et al. (\cite{deMink1}) who found in a sample of eclipsing SMC binaries, that initially wider systems, where spin-up was likely more efficient, seem more consistent with less conservative mass transfer models. 

Another puzzling result is the fact that the post-RLOF mass-ratio
deviates most from unity for the shortest period system (LZ\,Cep).
Therefore, in this system, the RLOF must have been comparatively more
efficient than in the longer period systems. As discussed above, the
mass transfer in LZ\,Cep was most likely highly non-conservative.
If the material lost by the system does not carry away angular momentum
in addition to the intrinsic orbital angular momentum, such a non-conservative
mass loss leads to a widening of the orbital separation $a$ given
by

\begin{center}
\[
\frac{\dot{a}}{a}=-\frac{\dot{M}}{M,}
\]

\par\end{center}

where $\dot{M}<0$ is the mass-loss rate (Singh \& Chaubey \cite{SC},
Tout \& Hall \cite{TH}). Assuming a fully non-conservative mass loss,
this would imply that the initial orbital period of LZ\,Cep must
have been shorter than 1.5\,days. The duration of the RLOF episode
depends on the evolution of the Roche-lobe filling factor of the mass
donor. The evolution of the size of the Roche lobe is set by the relative
importance of the orbit expansion and the shrinking of the relative
size of the mass donor's Roche lobe. In addition, the evolution of
the radius of the mass donor needs to be accounted for (see Tout \&
Hall \cite{TH}). Since massive stars with radiative envelopes contract
considerably on a dynamical timescale (Tout \& Hall \cite{TH}) and
in view of the current mass-ratio of LZ\,Cep (Mahy et al.\ \cite{Mahy1}),
it seems unlikely that a non-conservative RLOF could be the explanation
for a more efficient mass loss. Alternatively, if the material lost
by the system carried off additional angular momentum, then the result
could be a shrinking of the orbit. In active, low-mass stars, such
a situation could arise if the material lost by the donor corotates
with the latter under the effect of a magnetic field out to a wide
distance (Tout \& Hall \cite{TH}). However, strong magnetic fields
are fairly rare among single massive stars ($\sim7$\,\%), and their
incidence appears even lower among massive binaries (Neiner et al.\ \cite{Neiner}).
Hence, this scenario also appears very unlikely in the case of LZ\,Cep.

Clearly, the above trends call for confirmation by studying a larger
sample of post-case A RLOF massive binary systems. We are currently
undertaking such studies, and the results will be presented in forthcoming
publications.

\medskip{}

\subsection{Summary and conclusion}

We analysed the fundamental properties of the binary system HD 149\,404
by means of spectral disentangling and the CMFGEN atmosphere code.
Our investigation of this object represents a significant improvement
over previous studies because we have obtained the individual spectra
of each component. This led to the determination of a number of stellar
parameters that were used to constrain the evolutionary status of
the system.

We established the existence of a large overabundance in N and a C
and O depletion in the secondary star and also found a slight N enhancement
in the primary's spectrum. We showed that these surface abundances
cannot be explained by single-star evolutionary models. Furthermore,
we inferred an asynchronous rotation of the two stars of the system.
These two results indicate a previous mass and angular momentum-exchange
phase through a Roche-lobe overflow episode. Given the present status
of HD 149\,404 as an O+O binary, this mass transfer probably was
a case A RLOF. However, dedicated theoretical studies are needed to
understand the details of the evolution of this system. 
\begin{acknowledgements}
The Liège team acknowledges support from the Fonds de Recherche Scientifique
(FRS/FNRS) including especially an FRS/FNRS Research Project (T.0100.15),
as well as through an ARC grant for Concerted Research Actions financed
by the French Community of Belgium (Wallonia-Brussels Federation),
and an XMM PRODEX contract (Belspo). Anthony Hervé is supported by
grant 14-02385S from GA \v{C}R.\end{acknowledgements}

\end{document}